\documentclass[12pt,fleqn]{ucithesis}

\usepackage{amsmath}
\usepackage{amssymb}
\usepackage{bm}
\usepackage{booktabs}
\usepackage{array}
\usepackage{graphicx}
\usepackage{circuitikz}
\usepackage{tikz}
\usetikzlibrary{positioning, shapes.geometric}
\usepackage{upgreek}
\usepackage{makecell} 
\usepackage{float}
\usepackage[utf8]{inputenc}
\usepackage{textgreek}
\usepackage{xcolor} 
\usepackage{listings}
\DeclareUnicodeCharacter{2212}{\ensuremath{-}}
\usepackage[backend=biber, style=ieee]{biblatex}
\usepackage{relsize}
\usepackage[titletoc]{appendix}

\usepackage{caption}
\usepackage{subcaption}  
\usepackage{multirow}
\usepackage{tabularx}

\usepackage[plainpages=false,pdfborder={0 0 0}]{hyperref}

\thesistitle{Machine Learning Assisted Design of mmWave Wireless Transceiver Circuits}

\documenttitle{Thesis}

\degreename{MASTER OF SCIENCE}

\degreefield{Electrical and Computer Engineering}

\authorname{Xuzhe Zhao}

\committeechair{Assistant Professor Hamidreza Aghasi}
\othercommitteemembers
{
  Assistant Professor Rahim Esfandyar-Pour\\
  Assistant Professor Yanning Shen
}

\degreeyear{2024}

\copyrightdeclaration
{
  {\copyright} {\Degreeyear} \Authorname
}

\dedications
{
 To my parents and my friends, for their unwavering support and encouragement.
}

\acknowledgments
{
First and foremost, I would like to express my deepest gratitude to my advisor, Professor Hamidreza Aghasi. Without his guidance and persistent support, this thesis would not have been possible. I also would like to extend my appreciation to Professor Rahim Esfandyar-Pour and Professor Yanning Shen for kindly agreeing to serve as my committee members.

In addition, I would like to thank Asal Mehradfar, Yue Niu, Sara Babakniya, and Mahdi Alesheikh for their significant contributions to this work.

Also, I feel incredibly fortunate to have been part of the HIE lab during my studies at UCI. The discussion during group meetings and the passion of other lab members have always inspired me to delve deeper into this field. 

Portions of Chapter 3 to 5 are adapted to form the paper that is currently under review as “AICircuit: A Multi-Level Dataset and Benchmark for AI-Driven Analog Integrated Circuit Design”, in the 38th Conference on Neural Information Processing Systems (NeurIPS 2024) Track on Datasets and Benchmarks. The co-authors listed in this conference paper are Asal Mehradfar, Yue Niu, Sara Babakniya, Mahdi Alesheikh, Hamidreza Aghasi, and Salman Avestimehr. Professor Hamidreza Aghasi and Professor Salman Avestimehr of USC directed this work. 
}

\thesisabstract
{
 As fifth-generation (5G) and upcoming sixth-generation (6G) communications exhibit tremendous demands in providing high data throughput with a relatively low latency, millimeter-wave (mmWave) technologies manifest themselves as the key enabling components to achieve the envisioned performance and tasks. In this context, mmWave integrated circuits (IC) have attracted significant research interests over the past few decades, ranging from individual block design to complex system design. However, the highly nonlinear properties and intricate trade-offs involved render the design of analog or RF circuits a complicated process. The rapid evolution of fabrication technology also results in an increasingly long time allocated in the design process due to more stringent requirements. In this thesis, 28-GHz transceiver circuits are first investigated with detailed schematics and associated performance metrics. In this case, two target systems comprising heterogeneous individual blocks are selected and demonstrated on both the transmitter and receiver sides. Subsequently, some conventional and large-scale machine learning (ML) approaches are integrated into the design pipeline of the chosen systems to predict circuit parameters based on desired specifications, thereby circumventing the typical time-consuming iterations found in traditional methods. Finally, some potential research directions are discussed from the perspectives of circuit design and ML algorithms. 
}

\hypersetup{
	pdftitle={\Thesistitle},
	pdfauthor={\Authorname},
	pdfsubject={\Degreefield},
}

\begin{document}

\preliminarypages

\chapter{Introduction}
\section{Background}
Since the first generation (1G) of mobile communication was implemented to transfer the voice signal between different users, tremendous technologies relevant to wireless communication have evolved incredibly over the past several decades. To cope with the significantly increasing demands for wireless connectivity, the promising fifth generation (5G) and upcoming sixth generation (6G) wireless networks are deployed to alleviate spectrum crowding at lower frequencies with higher data throughput and lower latency \cite{rangan_millimeter-wave_2014}. From a numerical point of view, 5G and 6G communications are envisioned to provide the capacity of a data rate greater than 1-10 Gb/s, a latency on the order of 1 ms or less, and a high volume of traffic density from the unlicensed available bandwidth up to a few hundred GHz, which facilitates three major application scenarios comprising enhanced mobile broadband (eMBB), massive machine-type communication (mMTC), and ultra-reliable low latency communication (URLLC) \cite{dahlman_5g_2014}. Compared to previous communication technologies, 5G and 6G utilize millimeter-wave (mmWave) frequencies due to abundant spectrum resources. In that regard, many countries have allocated several mmWave bands within the frequency range 2 (FR2) from 24.25 to 52.6 GHz, and the higher band named WR-08 that is designated for 6G communication systems from 90 to 140 GHz \cite{pang_28-ghz_2019} \cite{maktoomi_sub-terahertz_2022}. Despite these benefits shown above, some key challenges should also be taken into consideration to employ the mmWave frequencies. For example, the mmWave signal suffers from large free space loss stated by Friis’ transmission law (Eq.\,(1.1)):
\begin{equation}
   P_r = P_t G_t G_r \left( \frac{c}{2 \pi f_c d} \right)^2 
\end{equation}
in which the gains of the transmitting and receiving antennas are captured by \(G_{\text{t}}\) and \(G_{\text{r}}\), the power of the received signal is inversely proportional to the square of carrier frequency and the distance between two devices. Moreover, the shadowing issue coming from various outdoor materials and large processing power consumption further impede the deployment of mmWave systems \cite{rangan_millimeter-wave_2014} \cite{zhao_28_2013}. Given that scenario, some key-enabling technologies have been called for overcoming these difficulties and achieving the necessary performance in the mmWave frequency bands, which consists of all-spectrum access, massive multiple-input multiple-output (MIMO), ultra dense networking, and semiconductor technologies \cite{gupta_survey_2015} \cite{boccardi_five_2014}. 

As an indispensable and promising path to achieve high-speed wireless communications, silicon based integrated circuit has been validated by designing wideband amplifiers over 30 GHz utilizing the mainstream CMOS technologies \cite{doan_millimeter-wave_2005}. Since then, analog or radio frequency (RF) integrated circuits operating in mmWave or sub-terahertz frequency bands have attracted a lot of research interest. In terms of individual circuit block, a Q-band low-noise amplifier (LNA), a dual-band linear up-conversion mixer, a compact super-harmonic voltage-controlled oscillator (VCO), and a three-way Doherty power amplifier (PA) are presented for various mmWave applications and fabricated in 90-nm, 40-nm, 65-nm, and 45-nm CMOS processes \cite{huang_3050-ghz_2023} \cite{bae_2440_2022} \cite{moradi_7682_2024} \cite{zhang_millimeter-wave_2023}. Benefit from advanced transistor technologies and reduced size of passive components, antenna manifests itself as a pivotal entity to provide efficient power transmission and reception with the expected radiation pattern operating in the mmWave and sub-terahertz frequency bands \cite{alesheikh_electronically_2024} \cite{maktoomi_broadband_2023}. To cope with severe path loss in these high frequency bands, a systematic design methodology is explored to maximize power generation and transfer with the nonlinear device model for GHz and THz applications \cite{saadat_low-power_2015} \cite{aghasi_design_2016} \cite{aghasi_092-thz_2017}. Pertaining to system-level implementation based on complex analog or RF integrated circuit blocks, a spatial-orthogonal Amplitude-Shift Keying (ASK) transmitter and a high-power transmitter equipped with phase locking techniques are demonstrated to mitigate the power limitation in the terahertz range \cite{aghasi_power-efficient_2017} \cite{han_sige_2015}. In addition, some dedicated receiver architectures using nonlinear analog operators are exhibited in mmWave single-input single-output (SISO) and multiple-input multiple-output (MIMO) communication systems \cite{shirani_quantifying_2022} \cite{shirani_mimo_2022}. As the size of transistors scales down, some advanced devices and sophisticated transceivers are utilized for high-speed communications, pattern recognition, molecular spectroscopy, real-time imaging, local-area sensing, and high resolution applications by leveraging the merits of the mmWave or THz frequency bands \cite{naskas_wideband_2022} \cite{aghasi_smart_2016} \cite{aghasi_88-ghz_2018} \cite{mostajeran_fully_2019} \cite{aghasi_millimeter-wave_2020} \cite{liu_49-63_2023}.

\section{Motivation}
Despite the fact that analog and RF integrated circuits are ubiquitous in mmWave wireless communication, the corresponding design pipeline can still be regarded as the significantly time-consuming and complicated process, even for experienced circuit designers. In general, there are three main steps in pre-layout circuit design. Firstly, circuit designers need to select the appropriate topology to meet the desired performance metrics by leveraging their prior knowledge and experience. Then, the initial size of each component shown in chosen topology is calculated according to a large number of approximations and simplifications. Following that, the specific circuit parameters are confirmed to meet a group of target specifications by performing the simulations countless times \cite{wang_gcn-rl_2020}. As the complexity of realized functions keeps growing and CMOS technology continues scaling, designing analog and RF integrated circuits in mmWave frequency bands has become more arduous. Recently, machine learning techniques have been significantly developed by means of high-performance computing hardware, which entails an accessible solution to a large number of interdisciplinary research in diversified domains \cite{mina_review_2022}. Regrading this approach, the digital circuit benefits from the established automation process by employing mature VLSI Computer-Aided Design (CAD) tools equipped with pre-sized defined logic libraries \cite{elfadel_machine_2019}. In contrast, analog design automation tools are not developing at the same pace as their digital counterpart, since there are a broad range of circuit topologies and intricate trade-offs involved in a variety of performance specifications, which makes design rules more complicated when layout parasitics are taken into account \cite{krylov_learning_2023}. Therefore, it is pivotal to accelerate analog and RF design procedures via identifying and automating the time-consuming and resource-intensive parts in an accurate and efficient manner. In this thesis, some conventional and large-scale machine learning algorithms are leveraged to assist the design of wireless transceiver circuits at 28 GHz from block-level and system-level perspectives. 

\section{Organization of This Thesis}
The remainder of this thesis is organized as follows. Chapter 2 investigates the overview of recent growing research works in analog circuit design enhanced by diversified machine learning algorithms. Chapter 3 and 4 demonstrate the 28 GHz transmitter and receiver design respectively, which comprises the performance analysis and the corresponding ocean scripts implementation. Chapter 5 presents a workflow involving different machine learning algorithms to optimize the system-level mmWave circuit design supported by the benchmark dataset. In the end, Chapter 6 concludes and summaries this work with some potential research directions in the future.

\chapter{Literature Review}
The aforementioned content indicates the feasibility and necessity of automated design flows applied in the analog and RF integrated circuit domain. In this section, some related work and representative methods will be discussed and reviewed. 

Figure 2.1 outlines the different categories of prior automating techniques for the analog and RF circuit design. These prevalent methods can first be divided into knowledge-based and optimization-based approaches \cite{barros_analog_2010}. In terms of knowledge-based techniques, the prior domain knowledge of experienced circuit designers is transcribed into the corresponding equations and algorithms, which means the complete design plan for a certain circuit topology needs to be well-defined with the detailed sizing solution of each component to fulfill the desired specifications \cite{lberni_analog_2024}. As an interactive design system based on knowledge-based techniques, IDAC is capable of sizing more than 40 analog schematics by means of providing some design knowledge regarding the relationship between the different technologies and block specifications \cite{degrauwe_idac_1987}. However, as the functions for analog and RF circuits become more sophisticated, it is significantly difficult to obtain the hand-crafting equations to describe the circuit knowledge in the corresponding programs. In that case, another main category named \begin{figure}[H]
    \centering
    \includegraphics[width=0.7\linewidth]{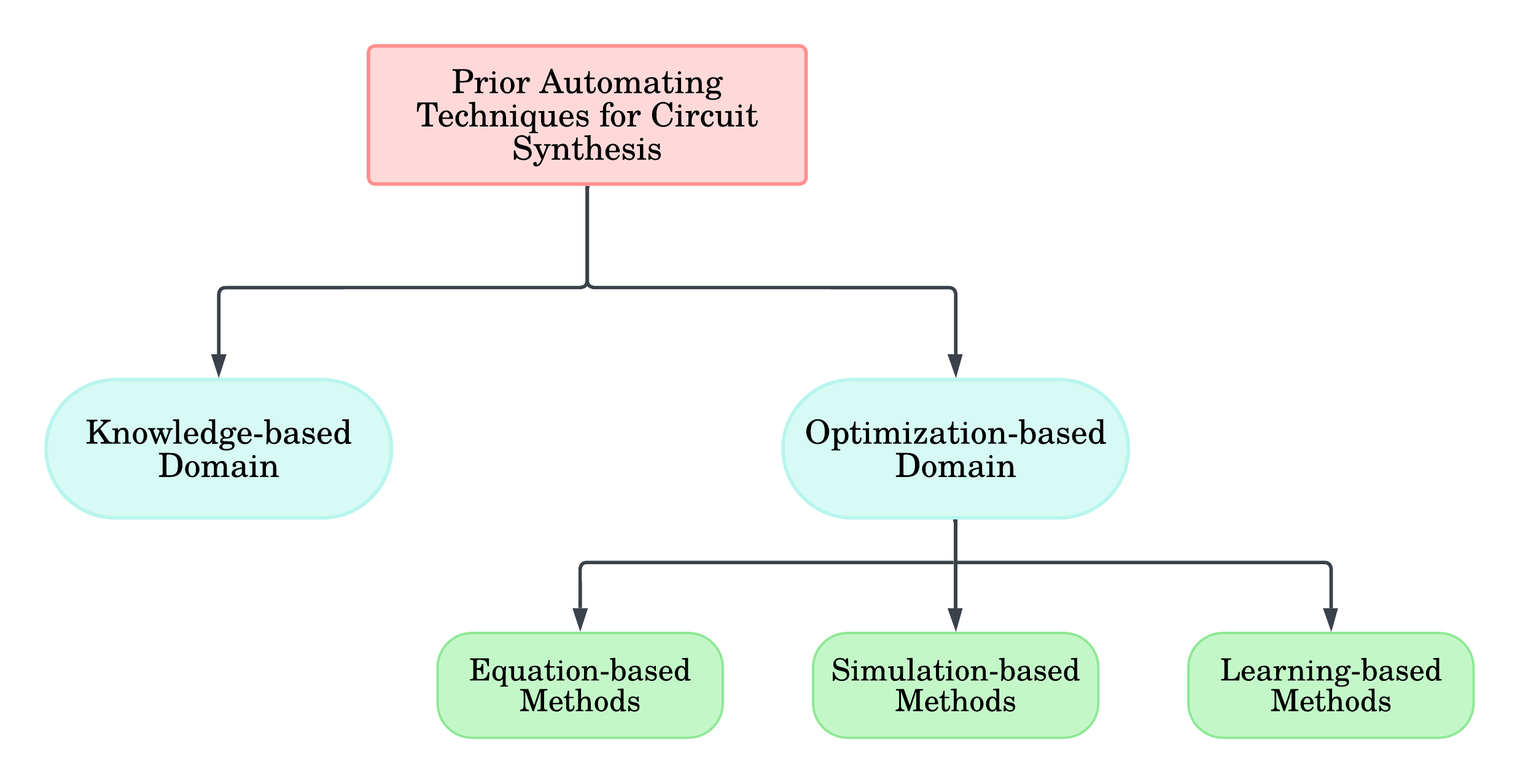}
    \caption{Classification for prior automating techniques for circuit synthesis.}
    \label{fig:Ch2_1}
\end{figure} optimization-based approach will be introduced. Rather than the need for a completed design plan, the design parameters and variables are updated at each iteration in optimization process, which will be finalized when the desired performance metrics are fulfilled \cite{barros_analog_2010}. According to different engines for optimization and evaluation, optimization-based technique can be further classified into three subsections including the equation-based, simulation-based, and learning-based methods as depicted in Figure 2.1. From the perspective of equation-based methods, the analytic constraint equations can be obtained manually or automatically with the aid of geometric programming and this circuit problem can be optimized using a numerical algorithm, such as GPCAD and TAGUS \cite{hershenson_optimal_2001} \cite{horta_analogue_2002}. The main issue for this method is substantially time-consuming to derive the appropriate equations and difficult to update them with the latest technology. Another popular approach used in analog circuit design optimization is simulation-based method. Typically, an initial population is selected following the stochastic sampling. After mutating the best children, the corresponding offspring will be examined and sampled from a new iteration. In the inner loop of  the optimization cycle, a circuit simulator tool, such as SPICE, is leveraged to determine and evaluate the performance of various circuits \cite{settaluri_autockt_2020}. Some representative works are presented through the implementation of Bayesian Optimization (BO), Evolutionary Algorithms (EA), and Particle Swarm Optimization (PSO) \cite{lyu_efficient_2018} \cite{phelps_anaconda_2000} \cite{prajapati_two_2015}. Based on the above investigation, this simulation-based approaches lead to higher accuracy since they conduct the optimization by directly collecting the simulated data. However, this method is subject to the convergence issue due to insufficient samples. Moreover, the entire optimization process needs to start over even though there is a slight change in desired performance specifications. With the significant development of relevant techniques, learning-based approaches obtain plenty of attraction in analog circuit domain. They primarily rely on the various machine learning methods to accurately establish the relationship between the circuit parameters and target performance metrics, and further solve the analog circuit problems, which entails the inverse process compared to the circuit simulator. Some supervised and reinforcement learning algorithms are used in prior works \cite{settaluri_autockt_2020} \cite{lourenco_exploration_2018} \cite{wang_gcn-rl_2020}.

In addition to the top-level overview for diversified automation techniques, the in-depth review for some innovative works is investigated in the following part. As depicted in Figure 2.2, the main methods utilized in these works come from supervised learning, reinforcement learning, and hybrid approaches combining neural network and optimization algorithms. \\

\begin{figure}[H]
    \centering
    \includegraphics[width=0.7\linewidth]{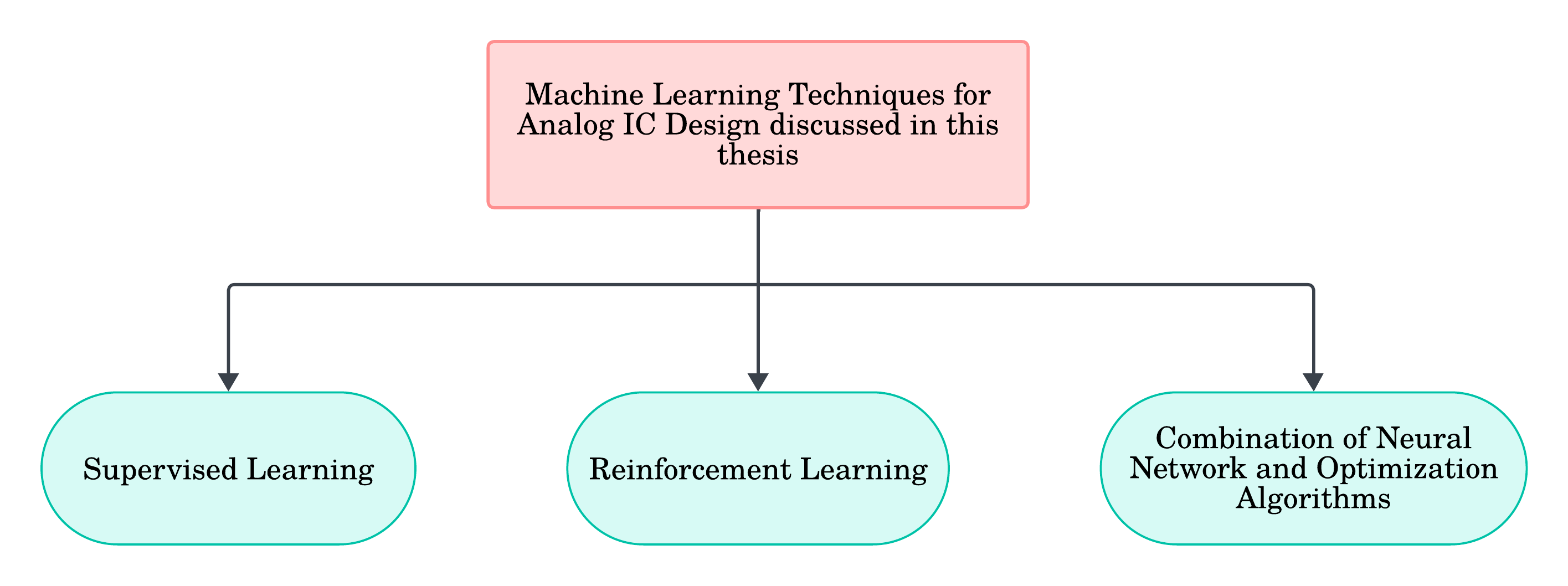}
    \caption{Machine learning techniques for analog circuit design reviewed in this thesis.}
    \label{fig:Ch2_2}
\end{figure}

\section{Supervised Learning}
As a typical and indispensable technique in the supervised learning domain, Neural Networks (NN) are attracting great attention in both industry and academia, such as speech recognition, imaging process, and natural language processing \cite{afacan_review_2021}. Inspired by human brain activity, NN have the fundamental linear threshold units whose name is perceptron to receive the input signal processed by proper weights and bias, then a specific activation function will be applied to deal with nonlinear problems. In general, there are two steps defined by the training and testing phases during the establishment of NN. In the first place, the real solution for different types of problem is approximated based on the labeled dataset. After that, some new outputs are accurately predicted for unknown data points as supported by the trained NN model \cite{aggarwal_neural_2018}.

From the perspective of analog and RF circuit design, NN is employed to estimate the functionality of different circuit topologies, which involves the relationship between the circuit parameters and performance metrics. Since Glenn Wolfe and Ranga Vemuri demonstrated the NN can be successfully used to assist the design of various CMOS operational amplifier (op-amp) in 2003, there are tremendous research works shown in analog circuit design based on the NN and its variants \cite{wolfe_extraction_2003}. According to \cite{wang_application_2018}, the Deep Neural Network (DNN) and the Recurrent Nural Network (RNN) are applied to solve the two-stage op-amp sizing problem. With appropriate sweeping for the aspect ratio of each MOSFET instance, the DNN and RNN models can achieve the accuracy of 95.6\% and 92.6\% for the 900 test circuit, which concludes that the DNN model has more stability than the RNN model attributed to less dependence of training data. Another disruptive work relies on the Artificial Neural Network (ANN) to predict the analog amplifier with voltage combiners that are utilized to boost the voltage gain. During the training phases, three different ANN models are trained with the same dataset. In the end, the model with 500 epochs has the best performance since it benefits from the value of weights shown in the ANN model validated through the entire dataset. Another contribution for this work is that the specification trade-offs are taken into consideration, which indicates the trained ANN model is capable of predicting the performance metrics outside the original training dataset with 0.0124 mean squared error \cite{lourenco_exploration_2018}.

\section{Reinforcement Learning}
As a pathway to access the general forms of artificial intelligence, reinforcement
learning (RL) is an instrumental technique inspired by human learning mechanisms \cite{aggarwal_neural_2018}. Compared to the supervised learning method, RL is driven by different rewards and penalties and does not require the collection of labeled datasets. In Figure 2.3, a general RL learning loop is presented. In each iteration, an RL agent is deployed to convert the current state (St) to a new action (At) given the corresponding reward (Rt). After that, the environment is employed to transform the current action (At) to a new state (St+1) and reward (Rt+1). This sequence of states and actions will lead to the maximum rewards ultimately. With this tight interaction between the agent and the environment, RL becomes prevalent in game-playing, self-driving cars, and intelligent robots \cite{zhao_deep_2020}.

\begin{figure}[H]
    \centering
    \vspace{10pt}
    \includegraphics[width=0.6\linewidth]{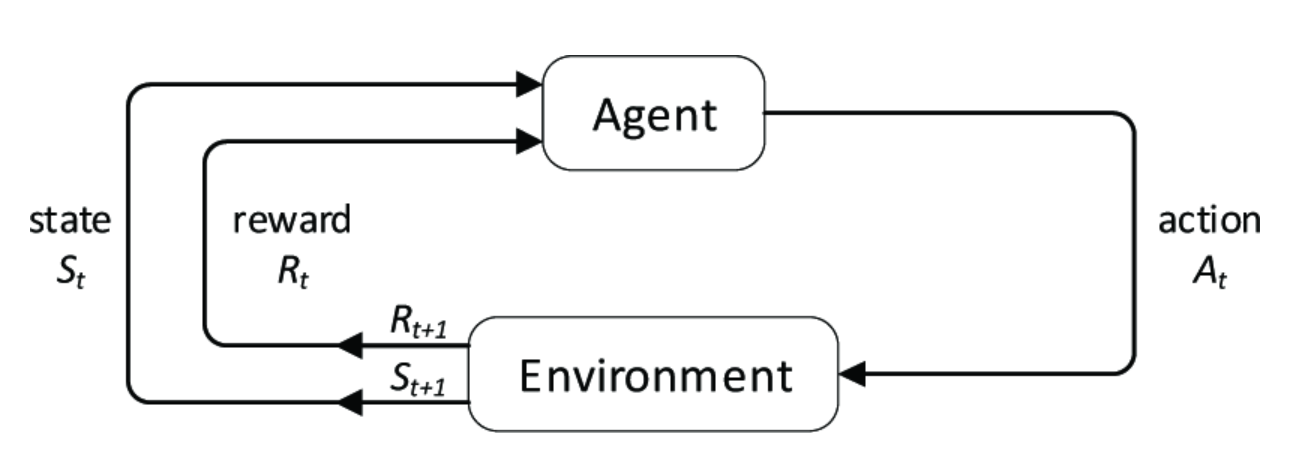}
    \caption{General learning loop for RL \cite{zhao_deep_2020}.}
    \label{fig:Ch2_3}
\end{figure}

In light of analog circuit design, RL agent is initialized by the performance metrics that can be regarded as the state of a circuit. Then, the action generated by the agent will be injected into the circuit simulator served as the environment. After comparing the predicted performance features with the desired specifications, the rewards will be computed and carry out the new state when their value increases. In contrast, the revised action will be generated due to lower rewards. The working flow for this process is shown in Figure 2.4 \cite{mina_review_2022}. \\

\begin{figure}[H]
    \centering
    \includegraphics[width=0.7\linewidth]{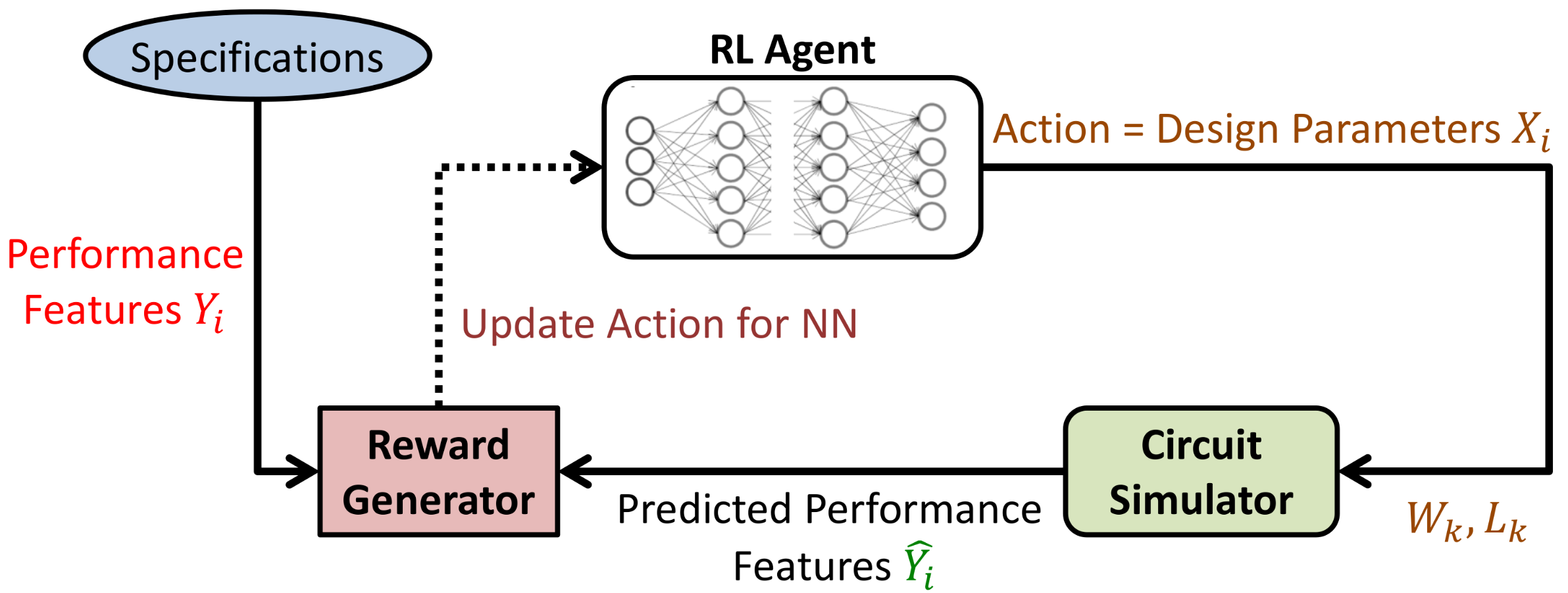}
    \caption{Design flow for RL used in analog circuit design \cite{mina_review_2022}.}
    \label{fig:Ch2_4}
\end{figure}

One representative work presented by Keertana Settaluri et al. shows a machine learning optimization framework that employes deep reinforcement learning can be deployed to design the transimpedance amplifier and two-stage op-amp \cite{settaluri_autockt_2020}. Aiming to improve the design efficiency and learning generalization, a sparse subsampling technique is utilized in this work to obtain the knowledge about the entire design space. Therefore, average 40× more sample efficient demonstrated here compared to prior optimization approaches. Another main contribution for this work is that they also take the layout parasitics into consideration by leveraging transfer learning, which mimics the more realistic design scenarios. For the experimental results of layout parasitics prediction, the two-stage op-amp with negative gm load is used to validate this AutoCkt model, which entails this framework has an ability to design 40 candidates that pass the LVS rules in under 3 days \cite{settaluri_autockt_2020}. In line with the same target to improve the sample efficiency, another ML model comprising the RL and the sensitivity analysis was proposed in 2023 \cite{choi_reinforcement_2023}. This work relies on the universal value function approximator (UVFA) in RL model training to convert the design space with high dimensions to one dimension of Figure of Merits (FoM) covering all performance specifications \cite{tom_schaul_universal_2015}. Apart from the method used in the RL model, a novel sampling method for data selection is described with reference to the 'Pick sampling' and 'PZ sampling' criteria. Furthermore, $\text{g}_\text{m}$/$\text{I}_\text{D}$ sizing methodology is applied to mitigate the short channel effect and alleviate the nonlinearity for prediction tasks. In the end, this proposed method is capable of optimizing the analog circuit sizing problems by 42.2\% improvement with the respect to convergence presented in a baseline work \cite{choi_reinforcement_2023} \cite{lillicrap_continuous_2019}.

\section{Combination of NN and Optimization Algorithms}
In addition to the supervised learning and reinforcement learning approaches articulated above, some automation techniques in analog circuit design are adopted to deploy the NN coupled with the various optimization algorithms. The general purpose for this combination is to benefit from both the efficiency and accuracy shown in NN and optimization algorithms respectively. For instance, there is a two-model chain implemented to predict the optimal performance under the new contexts including the loads and supply voltage, along with the trade-offs between them \cite{lourenco_using_2019}. In that work, a multi-variate polynomial regression is deployed to estimate the performance metrics and their trade-offs in the first layer. Following that, an ANN is invoked to find out the component sizing according to the desired specifications. Since data from the optimized sizing solution are used in the training stage, this framework can provide near-optimal predictions instantly even if the boundaries of desired performance are expanded, which is validated by sizing the folded cascade amplifier \cite{lourenco_using_2019}. Another strategy to incorporate the NN and different optimization algorithms is to accelerate the process of global optimization through replacing the circuit simulator by well-trained NN model. In that case, a fully-automated analog circuit generator framework named AnGeL is presented to cover all important steps in analog circuit schematic design involving the circuit topology determination and the circuit components sizing \cite{fayazi_angel_2023}. To cope with the entire design space and diversified potential topologies, NN is applied to estimate the functionality for more complicated circuit topologies by leveraging the knowledge from the simpler ones. During this phase, the complex circuit is divided into several small sub-sections, in which the less design parameters and smaller labeled training data are taken into consideration to obtain the optimum points in a fraction of the time. Moreover, these sub-circuits can be investigated in parallel as a result that this framework can support various circuit topologies efficiently. The interaction between the complicated topologies and the corresponding simpler sub-circuits is depicted in Figure 2.5. As a population-based approach, Particle Swarm Optimization (PSO) is also utilized in this work to enhance accuracy during the local optimization procedure. In the end, the experimental results demonstrate that AnGeL is capable of retaining the same accuracy but 2.9× to 75× faster than the state-of-the-art works, whereby testing more than 1450 different circuits \cite{fayazi_angel_2023}.

\captionsetup[figure]{justification=centering}
\begin{figure}[H]
    \centering
    \includegraphics[width=0.7\linewidth]{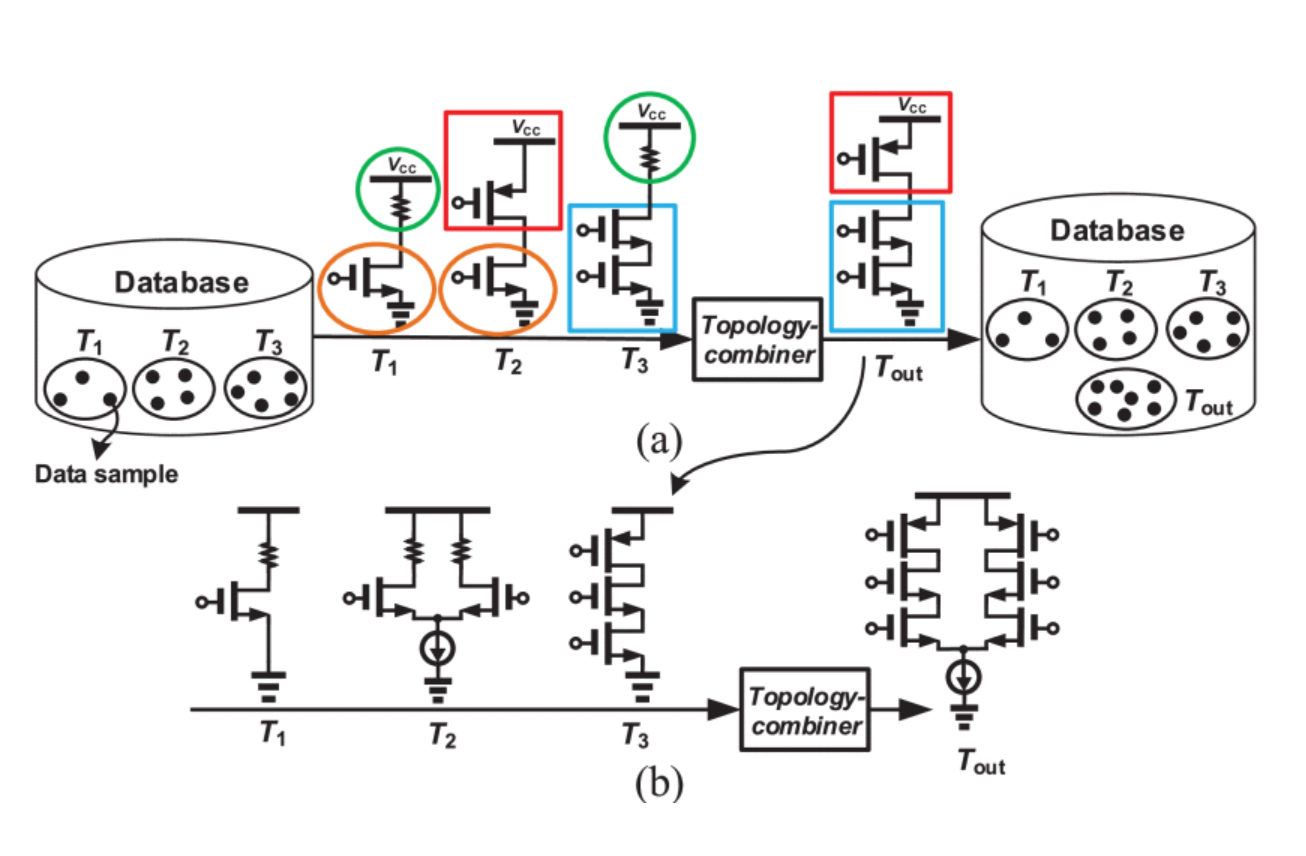}
    \caption{An example of leveraging the knowledge from (a) simpler circuits to (b) more complex one \cite{fayazi_angel_2023}.}
    \label{fig:Ch2_5}
\end{figure}
\chapter{A 28-GHz Transmitter Design}
Since the market demands are increasing for high data rate systems, mmWave communication shows the reliable capability to enhance the data rate to over 1 Gb/s. However, there are some inherent risks for electromagnetic waves propagating in such a high frequency band, which primarily comes from the large path loss and shadowing effect \cite{kim_28ghz_2019}. Aiming to overcome these detrimental effects, an RF transmitter is envisioned to deliver the mmWave signal with higher gain and output power while not degrading efficiency. For the sake of regulation, some indispensable functions and specifications are defined by different wireless standards for transceiver design, which ensures that the information can be successfully exchanged among various types of devices. As these requirements become more stringent in mmWave frequency, machine learning assisted methods are applied to facilitate the transmitter design. The target circuits and their desired performance metrics are presented in this section with the corresponding ocean scripts implementation. 

\section{Transmitter Architecture}
In this work, the direct conversion architecture is deployed in the transmitter design due to its higher linearity and lower power consumption, which eliminates the need for additional structure to convert Intermediate Frequency (IF) to Radio Frequency (RF) \cite{kim_28-ghz_2018}. With the intention of covering diversified circuit blocks in system-level design assisted by the ML approach, the combination of Voltage-Controlled Oscillator (VCO) and Power Amplifier (PA) is presented as a typical 'signal generator - amplifier' system on the transmitter side. As illustrated in Figure 3.1, the sustainable signal is first generated by the VCO with a tunable frequency controlled by an external voltage signal across a certain range. Then, this periodic signal is amplified via the PA with considerable gain and output power. Finally, the transmitting antenna allows this amplified signal to propagate into free space with a relatively long distance. Pertaining to the structure of individual blocks, the LC cross-coupled configuration is utilized in the VCO design to generate the oscillated signal by leveraging the DC supply power to counteract the parasitic losses in the LC resonator, as shown in Figure 3.2 and 3.3. Owing to its low phase noise and large output swing, this type of VCO has become popular in high-speed wireless communication systems \cite{buonomo_finding_2008}. For the sake of system stability, a source-follower buffer is placed following the VCO to maintain the continuous oscillation.  As the most power-hungry building block in the transmitter, PA is implemented by the two-stage differential cascode structure illustrated in Figure 3.4 and 3.5 as a result to amplify the signal with high power gain without compromising efficiency \cite{abbasi_broadband_2010}.

\begin{figure}[!htb]
    \centering
        \begin{circuitikz}[line width=0.2mm,scale=0.8, transform shape]
        \draw (0,0) node[oscillator,anchor=south](VCO){}; 
        \draw (VCO.north) node[above=4pt] {\parbox{5cm}{\centering $\text{Voltage-Controlled}$ \\ \vspace{1pt} $\text{Oscillator}$}};

        \draw (VCO.east) to[amp,name=buffer] ++(3.2,0);
        \draw (buffer.west) node[inputarrow,scale=1.5]{};
        \draw (buffer.south) node[below=5pt]{$\text{Buffer}$};
        \draw ([xshift=0.2cm,yshift=0.1cm]VCO.center) node[rectangle,draw,densely dashed,minimum width=5.5cm, minimum height=3.5cm](box){};
        \draw (box.center) node[below=35pt] {$\texttt{28\,GHz Signal Generator}$};

        \draw (buffer.east) to[amp,name=PA1] ++(5,0);
        \draw (PA1.west) node[inputarrow,scale=1.5]{};
        \draw (PA1.east) to[amp,name=PA2,xshift=-1.5cm,fill=white] ++(2.5,0);
        \draw (PA2.south) node[below=5pt]{$\text{Power Amplifier}$};

        \draw ([xshift=0.4cm]PA2.east) node[txantenna,anchor=center](Txantenna){};
        \draw ([xshift=0.5cm]Txantenna.north) node[above=10pt] {$\text{RF Signal}$};
    \end{circuitikz}
    \caption{A 28\,GHz direct conversion transmitter architecture involving VCO and PA.}
    \label{fig:Tx}
\end{figure}
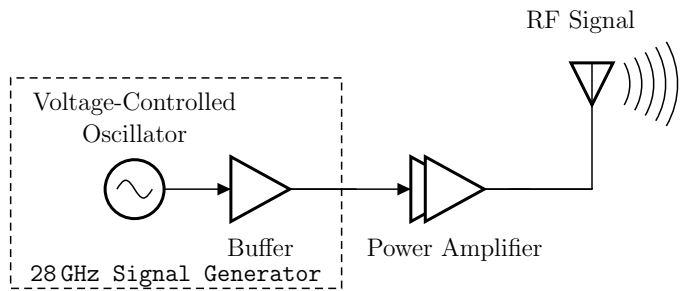
\begin{figure}[!htb]
    \centering
        \begin{circuitikz}[line width=0.1mm,scale=0.8, transform shape]
        \ctikzset{tripoles/mos style/arrows}
        \ctikzset{capacitors/scale=0.5}
        \ctikzset{resistors/scale=0.5}
        \ctikzset{inductors/scale=0.7, inductors/coils=4}
        \ctikzset{transistors/scale=0.8}
        \ctikzset{grounds/scale=0.8}
        \ctikzset{sources/scale=0.7}

        \draw (0,0) node[nmos,anchor=G,xscale=-1](M1){\ctikzflipx{$\text{W}_\text{N1}$}};
        \draw ([yshift=0.8cm]M1.D) to[short,-o] ($(M1.D)+(-1.5,0.8)$) node[left] {$\text{V}_\text{out}\text{+}$};
        
        \draw (1.5,0) node[nmos,anchor=G](M2){$\text{W}_\text{N1}$};
        \draw ([yshift=0.8cm]M2.D) to[short,-o] ($(M2.D)+(1.5,0.8)$) node[right] {$\text{V}_\text{out}\text{-}$};

        \draw ([xshift=-1.6cm,yshift=0.4cm]M1.D) node[nmos,anchor=G,rotate=-90](M3){};
        \draw ([xshift=-0.2cm]M3.S) node[anchor=south] {$\text{W}_\text{var}$};
        \draw (M3.G) -- ([xshift=1.6cm]M3.G);
        \draw (M3.D) -- ([yshift=-0.4cm]M3.D);
        \draw (M3.S) -- ([yshift=-0.4cm]M3.S);
        \draw ([yshift=-0.4cm]M3.D) -- ([yshift=-0.4cm]M3.S);

        \draw ([xshift=1.6cm,yshift=0.4cm]M2.D) node[nmos,anchor=G,rotate=-90](M4){};
        \draw ([xshift=0.2cm]M4.D) node[anchor=south] {$\text{W}_\text{var}$};
        \draw (M4.G) -- ([xshift=-1.6cm]M4.G);
        \draw (M4.D) -- ([yshift=-0.4cm]M4.D);
        \draw (M4.S) -- ([yshift=-0.4cm]M4.S);
        \draw ([yshift=-0.4cm]M4.D) -- ([yshift=-0.4cm]M4.S);
        \draw ([yshift=-1.2cm]M3.G) -- ([yshift=-2cm]M3.G);
        \draw ([yshift=-1.2cm]M4.G) -- ([yshift=-2cm]M4.G);
        \draw ([yshift=-2cm]M3.G) -- ([yshift=-2cm]M4.G);
        \draw ([xshift=1cm,yshift=-2cm]M3.G) to[short,-o] ($(M3.G)+(1,-2.4)$) node[below] {$\text{V}_\text{control}$};

        \draw (M1.S) -- (M2.S);
        \draw ([xshift=1.55cm]M1.S) -- ([xshift=1.55cm,yshift=-0.8cm]M1.S);
        \draw ([xshift=1.55cm,yshift=-0.8cm]M1.S) node[nmos,anchor=D](M5){$\text{W}_\text{N3}$};

        \draw ([xshift=-3.6cm]M5.G) node[nmos,anchor=G,xscale=-1](M6){\ctikzflipx{$\text{W}_\text{N2}$}};
        \draw (M6.G) -- (M5.G);

        \draw ([yshift=2cm]M6.D) to[american current source,name=I1,invert] ++(0,2);
        \draw ([yshift=0.7cm]I1.left) -- ([yshift=1.4cm]I1.left);
        \draw (M6.D) -- ([yshift=2cm]M6.D);
        \draw ([yshift=0.5cm]M6.D) -- ([xshift=1.5cm,yshift=0.5cm]M6.D);
        \draw ([xshift=1.5cm,yshift=0.5cm]M6.D) -- ([xshift=1.5cm,yshift=-0.6cm]M6.D);
        
        \draw (M1.G) -- ([xshift=0.2cm]M1.G);
        \draw (M2.G) -- ([xshift=-0.2cm]M2.G);
        \draw ([yshift=0.1cm]M1.D) -- ([xshift=0.8cm,yshift=0.1cm]M1.D);
        \draw ([yshift=0.1cm]M2.D) -- ([xshift=-0.8cm,yshift=0.1cm]M2.D);
        \draw ([xshift=0.2cm]M1.G) -- ([xshift=-0.8cm,yshift=0.1cm]M2.D);
        \draw ([xshift=-0.2cm]M2.G) -- ([xshift=0.8cm,yshift=0.1cm]M1.D);

        \draw (M1.D) -- ([yshift=1.2cm]M1.D);
        \draw ([yshift=1.2cm]M1.D) node[circ]{};
        \draw ([yshift=1.2cm]M1.D) to[R,l=$\text{R}_\text{p}$,name=R1] ++(0,1.5);
        \draw (M2.D) -- ([yshift=1.2cm]M2.D);
        \draw ([yshift=1.2cm]M2.D) node[circ]{};
        \draw ([yshift=1.2cm]M2.D) to[R,l=$\text{R}_\text{p}$,name=R2] ++(0,1.5);

        \draw ([xshift=-0.8cm,yshift=-0.75cm]R1.north) to[C,l=$\text{C}$,name=C1] ++(0,1.5);
        \draw ([yshift=-0.67cm]C1.west) -- ([xshift=0.9cm,yshift=-0.67cm]C1.west);
        \draw ([xshift=0.8cm,yshift=-0.75cm]R2.south) to[C,l=$\text{C}$,name=C2] ++(0,1.5);
        \draw ([yshift=-0.67cm]C2.west) -- ([xshift=-0.9cm,yshift=-0.67cm]C2.west);

        \draw ([xshift=0.8cm,yshift=-0.75cm]R1.south) to[L,l=$\text{L}$,name=L1] ++(0,1.5);
        \draw ([yshift=-0.45cm]L1.west) -- ([xshift=-0.9cm,yshift=-0.45cm]L1.west);
        \draw ([xshift=-0.8cm,yshift=-0.75cm]R2.north) to[L,l=$\text{L}$,name=L2] ++(0,1.5);
        \draw ([yshift=-0.45cm]L2.west) -- ([xshift=0.9cm,yshift=-0.45cm]L2.west);

        \draw [line width=0.5mm]([yshift=0.75cm,xshift=-4.2cm]R1.north)--([yshift=0.75cm,xshift=2cm]R2.north) node[above,near end]{$\textbf{V}_\textbf{DD}$};

        \draw (M5.S) node[ground](GND){};
        \draw (M6.S) node[ground](GND){};

    \end{circuitikz}
    \caption{LC cross-coupled VCO schematic.}
    \label{fig:VCO}
\end{figure}
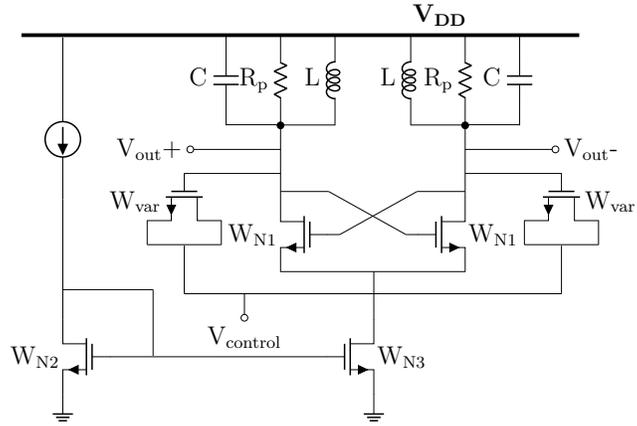
\begin{figure}[!htb]
    \centering
    \vspace{20pt}
    \includegraphics[width=0.8\textwidth]{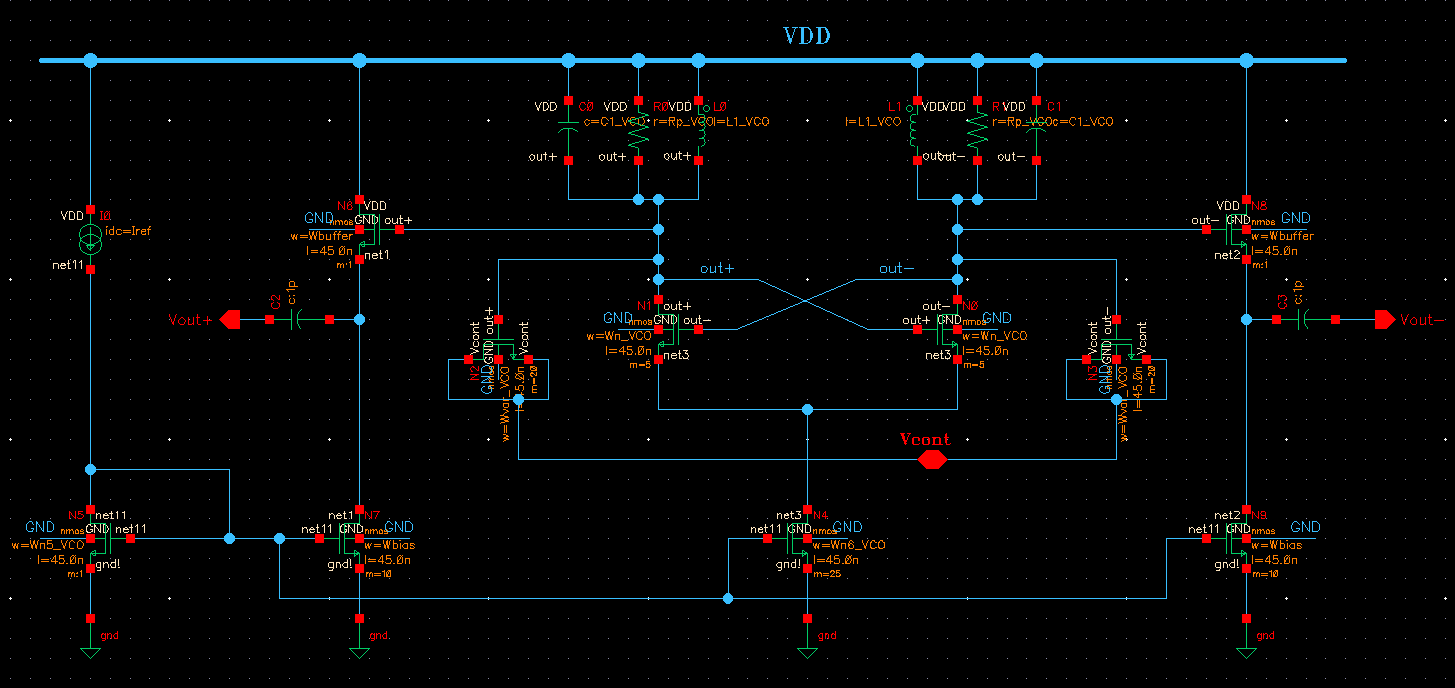}
    \caption{VCO implementation with buffer using NCSU 45\,nm PDK in Cadence Virtuoso}
    \label{fig:VCO_Cadence}
    \vspace{20pt}
\end{figure}
\begin{figure}[t]
    \centering
        \begin{circuitikz}[line width=0.1mm,scale=0.8, transform shape]
        \ctikzset{tripoles/mos style/arrows}
        \ctikzset{resistors/scale=0.5}
        \ctikzset{inductors/scale=0.8, inductors/coils=4}
        \ctikzset{transistors/scale=0.8}
        \ctikzset{quadpoles/transformer/inner=1, quadpoles/transformer/width=1}

        \draw (0,0) node[nmos,anchor=G,rotate=-90](M1){};
        \draw ([yshift=-1.4cm]M1.G) node[anchor=south] {$\text{W}_\text{N1}$};

        \draw (2.2,0) node[nmos,anchor=G,rotate=-90](M2){};
        \draw ([yshift=-1.4cm]M2.G) node[anchor=south] {$\text{W}_\text{N1}$};
        \draw (M2.G) to[short,-*] ($(M2.G)+(0,0.1)$) node[above] {$\text{V}_\text{b1}$};

        \draw (0,-3) node[nmos,anchor=G,rotate=90,yscale=-1](M3){};
        \draw ([yshift=0.8cm]M3.G) node[anchor=south] {$\text{W}_\text{N1}$};

        \draw (2.2,-3) node[nmos,anchor=G,rotate=90,yscale=-1](M4){};
        \draw ([yshift=0.8cm]M4.G) node[anchor=south] {$\text{W}_\text{N1}$};
        \draw (M4.G) to[short,-*] ($(M4.G)+(0,-0.1)$) node[below] {$\text{V}_\text{b1}$};

        \draw (6.12,0) node[nmos,anchor=G,rotate=-90](M5){};
        \draw ([yshift=-1.4cm]M5.G) node[anchor=south] {$\text{W}_\text{N2}$};

        \draw (8.32,0) node[nmos,anchor=G,rotate=-90](M6){};
        \draw ([yshift=-1.4cm]M6.G) node[anchor=south] {$\text{W}_\text{N2}$};
        \draw (M6.G) to[short,-*] ($(M6.G)+(0,0.1)$) node[above] {$\text{V}_\text{b2}$};

        \draw (6.12,-3) node[nmos,anchor=G,rotate=90,yscale=-1](M7){};
        \draw ([yshift=0.8cm]M7.G) node[anchor=south] {$\text{W}_\text{N2}$};

        \draw (8.32,-3) node[nmos,anchor=G,rotate=90,yscale=-1](M8){};
        \draw ([yshift=0.8cm]M8.G) node[anchor=south] {$\text{W}_\text{N2}$};
        \draw (M8.G) to[short,-*] ($(M8.G)+(0,-0.1)$) node[below] {$\text{V}_\text{b2}$};

        \draw (M1.D) to[L,l=$\text{L}_\text{m}$,name=L1] ++(1,0);
        \draw (M3.D) to[L,l_=$\text{L}_\text{m}$,name=L1,mirror] ++(1,0);
        \draw (M5.D) to[L,l=$\text{L}_\text{m}$,name=L1] ++(1,0);
        \draw (M7.D) to[L,l_=$\text{L}_\text{m}$,name=L1,mirror] ++(1,0);
        
        \draw ([xshift=-1.4cm,yshift=0.15cm]M1.G) node[transformer,anchor=B1](T1){};
        \draw (T1.inner dot A1) node[circ]{};
        \draw (T1.inner dot B1) node[circ]{};
        \draw [>=latex, <->] (-2.35,0.1) to[bend left=45] (-1.55,0.1);
        \draw (T1.B1) -- ([xshift=1.4cm]T1.B1);
        \draw ([xshift=1.4cm]T1.B1) -- (M1.G);
        \draw ([xshift=0.05cm]T1-L1.midtap) node[anchor=east] {$\text{L}_\text{ip}$};
        \draw ([xshift=-0.05cm]T1-L2.midtap) node[anchor=west] {$\text{L}_\text{is}$};
        \draw (T1.B2) node[anchor=south,rotate=90] {$\text{V}_\text{b3}$};
        \draw (T1.A1) to[short,-o] ($(T1.A1)+(-1,0)$) node[left] {$\text{V}_\text{in}\text{+}$};

        \draw 
        (T1.B2) node[transformer,anchor=B1](T2){};
        \draw (T2.inner dot A2) node[circ]{};
        \draw (T2.inner dot B2) node[circ]{};
        \draw [>=latex, <->] (-2.35,-3.1) to[bend right=45] (-1.55,-3.1);
        \draw (T2.B2) -- ([xshift=1.4cm]T2.B2);
        \draw ([xshift=1.4cm]T2.B2) -- (M3.G);
        \draw ([xshift=0.05cm]T2-L1.midtap) node[anchor=east] {$\text{L}_\text{ip}$};
        \draw ([xshift=-0.05cm]T2-L2.midtap) node[anchor=west] {$\text{L}_\text{is}$};
        \draw (T2.A2) to[short,-o] ($(T2.A2)+(-1,0)$) node[left] {$\text{V}_\text{in}\text{-}$};

        \draw ([xshift=1.4cm,yshift=0.15cm]M2.G) node[transformer,anchor=A1](T3){};
        \draw (T3.inner dot A1) node[circ]{};
        \draw (T3.inner dot B1) node[circ]{};
        \draw [>=latex, <->] (3.76,0.1) to[bend left=45] (4.56,0.1);
        \draw (T3.A1) -- ([xshift=-0.78cm]T3.A1);
        \draw ([xshift=-0.78cm]T3.A1) -- ([xshift=-0.78cm,yshift=-0.94cm]T3.A1);
        \draw (T3.B1) -- ([xshift=1.4cm]T3.B1);
        \draw ([xshift=1.4cm]T3.B1) -- (M5.G);
        \draw (T3.A2) node[anchor=south,rotate=90] {$\text{V}_\text{DD}$};
        \draw (T3.B2) node[anchor=south,rotate=90] {$\text{V}_\text{b4}$};

        \draw 
        (T3.A2) node[transformer,anchor=A1](T4){};
        \draw (T4.inner dot A2) node[circ]{};
        \draw (T4.inner dot B2) node[circ]{};
        \draw [>=latex, <->] (3.76,-3.1) to[bend right=45] (4.56,-3.1);
        \draw (T4.A2) -- ([xshift=-0.78cm]T4.A2);
        \draw ([xshift=-0.78cm]T4.A2) -- ([xshift=-0.78cm,yshift=1cm]T4.A2);
        \draw (T4.B2) -- ([xshift=1.4cm]T4.B2);
        \draw ([xshift=1.4cm]T4.B2) -- (M7.G);

        \draw ([xshift=1.4cm,yshift=0.15cm]M6.G) node[transformer,anchor=A1](T5){};
        \draw (T5.inner dot A1) node[circ]{};
        \draw (T5.inner dot B1) node[circ]{};
        \draw [>=latex, <->] (9.88,0.1) to[bend left=45] (10.68,0.1);
        \draw (T5.A1) -- ([xshift=-0.78cm]T5.A1);
        \draw ([xshift=-0.78cm]T5.A1) -- ([xshift=-0.78cm,yshift=-0.94cm]T5.A1);
        \draw ([xshift=0.1cm]T5-L1.midtap) node[anchor=east] {$\text{L}_\text{op}$};
        \draw ([xshift=-0.05cm]T5-L2.midtap) node[anchor=west] {$\text{L}_\text{os}$};
        \draw (T5.A2) node[anchor=south,rotate=90] {$\text{V}_\text{DD}$};
        \draw (T5.B1) to[short,-o] ($(T5.B1)+(1,0)$) node[right] {$\text{V}_\text{out}\text{+}$};

        \draw 
        (T5.A2) node[transformer,anchor=A1](T6){};
        \draw (T6.inner dot A2) node[circ]{};
        \draw (T6.inner dot B2) node[circ]{};
        \draw [>=latex, <->] (9.88,-3.1) to[bend right=45] (10.68,-3.1);
        \draw (T6.A2) -- ([xshift=-0.78cm]T6.A2);
        \draw ([xshift=-0.78cm]T6.A2) -- ([xshift=-0.78cm,yshift=1cm]T6.A2);
        \draw ([xshift=0.1cm]T6-L1.midtap) node[anchor=east] {$\text{L}_\text{op}$};
        \draw ([xshift=-0.05cm]T6-L2.midtap) node[anchor=west] {$\text{L}_\text{os}$};
        \draw (T6.B2) to[short,-o] ($(T6.B2)+(1,0)$) node[right] {$\text{V}_\text{out}\text{-}$};

        \draw (M1.S) node[ground](GND){};
        \draw (M3.S) node[ground,yscale=-1](GND){};
        \draw (M5.S) node[ground](GND){};
        \draw (M7.S) node[ground,yscale=-1](GND){};

    \end{circuitikz}
    \caption{Two-stage differential cascode PA schematic.}
    \label{fig:PA}
\end{figure}
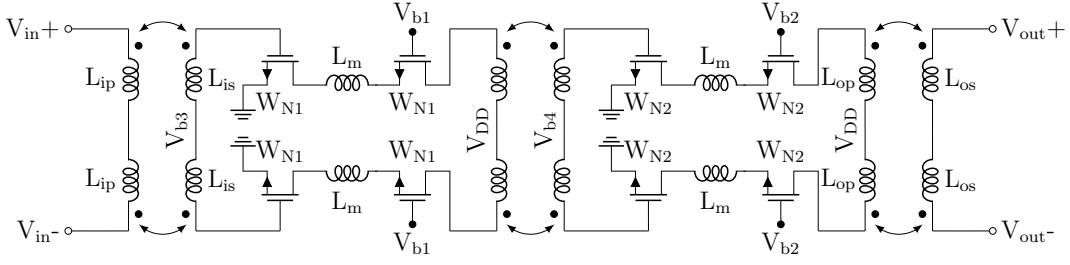
\begin{figure}[!htb]
    \centering
    \vspace{20pt}
    \includegraphics[width=0.9\textwidth]{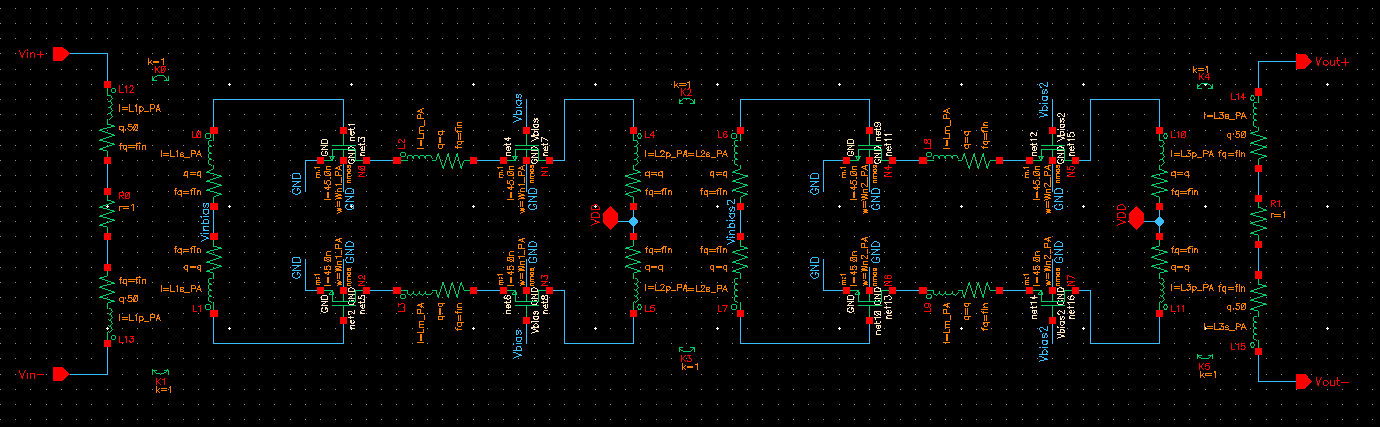}
    \caption{PA implementation using NCSU 45\,nm PDK in Cadence Virtuoso}
    \label{fig:PA_Cadence}
    \vspace{20pt}
\end{figure}

\section{Performance Specifications Analysis}
\subsection{Individual Performance}
In this section, some representative performance metrics are investigated for VCO and PA individually. These candidates are also used to train the ML model and further predict the related circuit parameters.

Since VCO is required to output sustainable signals whose frequency is varied across a certain range, the first important specification should be intended to quantify frequency variation, called the 'tuning range'. During the tuning process, MOS varactors are deployed to vary the resonance frequency in LC cross-coupled VCO by means of changing their capacitance \text{\(C_{\text{var2}}-C_{\text{var1}}\)} with different control voltage. According to an approximation of \(C_{\text{var}}<<C_1\), this specification can be expressed as Eq.\,(3.1):
\begin{equation}
   \Delta\omega \approx \frac{1}{\sqrt{L_1 C_1}} \frac{C_{\text{var2}} - C_{\text{var1}}}{2 C_1}
\end{equation}
Another essential characteristic is identified as the “phase noise”, indicating the small phase deviation due to the noise of different components in the VCO. This random departure is quantified by normalizing the noise power to the carrier power with the respect to specific offset frequency. The unit “dBc/Hz” also captures the relationship between these two types of power by integrating the in-band and out-of-band components as a result to further signify the signal-to-noise ratio (SNR) performance of the system with the VCO inside \cite{moradi_compact_2022}. In order to cover the multiple noise sources comprising the thermal noise from the LC tank and transistors, along with the flicker noise, Lesson’s model is leveraged to represent the noise effect through indicating the relationship between the high frequency noise floor and the flicker noise up-conversion, as shown in Eq.\,(3.2) \cite{lee_oscillator_2000}: 
\begin{equation}
   L\{\Delta\omega\} = 10 \log \left[ \frac{2 F k T}{P_{\text{sig}}} \left(1 + \left(\frac{\omega_0}{2 Q \Delta\omega}\right)^2\right) \left(1 + \frac{\Delta\omega_{1/f^3}}{|\Delta\omega|}\right) \right] \, (\text{dBc/Hz})
\end{equation}
Moreover, there is an intrinsic trade-off between the tuning range and the phase noise primarily coming from the different optimum scenarios regarding the capacitance ratio between the switching capacitors and tuning varactors, which leads to more complicated design procedure \cite{liu_single-switch_2024}. 

As a key enabling block to increase the power level of the output signal, there are several critical characteristics that should be taken into consideration during the design of PA. The first performance metric called the “large-signal power gain” is employed to describe the ratio of output power to input power when PA reaches the saturation region with increased input power. Given the great amount of power consumption by the PA, efficiency is also an important characteristic for evaluating power-intensive devices. The detailed assessment of PA efficiency is defined by two metrics, including “drain efficiency” (Eq.\,(3.3)) and “power-added efficiency (PAE)” (Eq.\,(3.4)):
\begin{equation}
  \text{Drain Efficiency}\,(\eta) = \frac{P_{\text{out}}}{P_{\text{DC}}}\times 100\% 
\end{equation}
\vspace{-1cm}
\begin{equation}
  \text{PAE} = \frac{P_{\text{out}} - P_{\text{in}}}{P_{\text{DC}}}\times 100\% 
\end{equation}
The former is utilized to measure the delivered output power with the respect to the DC power. Considering the power used by the amplifier itself, the PAE examines the efficiency for the RF power that is added to the device by subtracting the input RF power. In this work, both types of efficiency are investigated with the variation of circuit parameters.

\subsection{System Performance}
For the RF transmitter system design, the output power level is specified by a rule named “spectral mask”, in which the maximum transmission power is limited by a certain level and out-of-channel emissions are also required to decrease within the dedicated bandwidth, thereby reducing adjacent-channel interference. In that regard, the output power of the system combining VCO and PA is taken into account with the measurement of the 3\,dB bandwidth. Besides these two specifications, the dc power consumption is applied to assess the power utilized by the system during its operation. Finally, the voltage swing performance implies the ability of the system to deal with large input signals without distortion, which directly determines the working range of the transmitter. 

\section{Ocean Scripts Implementation}
In light of performing repetitive tasks in Cadence, conventional ADE simulations are not suitable for large-scale simulations, which is a relatively essential step for collecting the training data during the ML model training process. In that case, the new entity called OCEAN scripts is implemented to perform the parametric analysis and process the simulation data from the command line in the specified UNIX shell \cite{cadence_design_systems_inc_ocean_2004}. As a subset of the SKILL language, OCEAN abbreviated from Open Command Environment for Analysis is deployed to facilitate the various simulations and establish the interface between Cadence and the Python environment in this work.

By leveraging the ocean scripts and the defined schematic netlist, the various analysis and simulation results can be encapsulated into different commands. During the system design on the transmitter side, dc and transient analysis are employed to acquire the power consumption and voltage swing performance from the peak-to-peak value of the output waveform. As for the 3dB bandwidth, S Parameters (SP) analysis is used to select the frequency difference across the points decreased by 3dB with reference to the maximum amplitude. Another critical analysis commonly deployed in the RF circuit domain is periodic steady-state (PSS) analysis, in which the device under test is excited by a large periodic signal or the system has the periodic operating point. In that regard, the Harmonic Balance (HB) engine is applied to evaluate the system output power in the PSS analysis, which also includes the VCO performance comprising the phase noise and tuning range, and the PA specifications consisting of power gain, drain efficiency, and PAE. The system schematic in Cadence and its performance metrics with the ocean scripts implementation are demonstrated in Figure 3.6 and Table 3.1 respectively.

\begin{figure}[!htb]
    \centering
    \vspace{20pt}
    \includegraphics[width=1\textwidth]{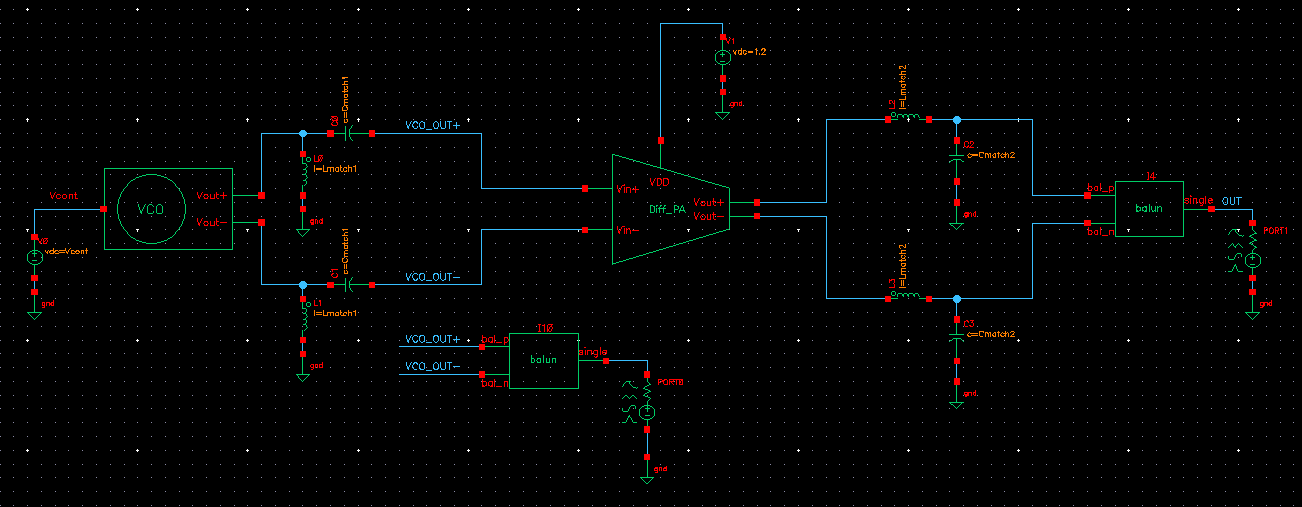}
    \caption{Schematic of target system on transmitter side in Cadence Virtuoso.}
    \label{fig:Tx_Cadence}
\end{figure}
\begin{table}[t]
    \centering
    \renewcommand{\arraystretch}{1.5}
    {\small \begin{tabular}{m{0.3\textwidth}m{0.6\textwidth}}
        \hline
        \textbf{Performance Metrics} & \textbf{Ocean Scripts}\\
        \hline\hline
         Power Consumption& \texttt{getData(":pwr" ?result "dcOp")}\\
         \hline
         Bandwidth& \texttt{(ymax(cross(db10(gt(sp(1 1 ?result "sp") sp(1 2 ?result "sp") sp(2 1 ?result "sp") sp(2 2 ?result "sp"))) (ymax(db10(gt(sp(1 1 ?result "sp") sp(1 2 ?result "sp") sp(2 1 ?result "sp") sp(2 2 ?result "sp")))) - 3) 2 "either" t "time" nil)) - ymin(cross(db10(gt(sp(1 1 ?result "sp") sp(1 2 ?result "sp") sp(2 1 ?result "sp") sp(2 2 ?result "sp"))) (ymax(db10(gt(sp(1 1 ?result "sp") sp(1 2 ?result "sp") sp(2 1 ?result "sp") sp(2 2 ?result "sp")))) - 3) 2 "either" t "time" nil)))}\\
         \hline
         Output Power& \texttt{ymax(dbm(pvr('pss "/OUT" "/gnd!" 50.0 '(1))))}\\
         \hline
         Voltage Swing& \texttt{ymax(v("/OUT" ?result "tran")) - ymin(v("/OUT" ?result "tran"))}\\
         \hline
         Tuning Range for VCO& \texttt{ymax(harmonic(xval(getData("/VCO\_OUT+" ?result "pss\_fd")) '1)) - ymin(harmonic(xval(getData("/VCO\_OUT+" ?result "pss\_fd")) '1))}\\
         \hline
         Phase Noise for VCO& \texttt{value(leafValue(pn('pnoise) "Vcont" 0.6) 1000000)}\\
         \hline
         Power Gain for PA& \texttt{ymax(db10((pvi('pss "/net7" "/net4" "/I19/Vout+" 0 '(1)) / (- pvi('pss "/VCO\_OUT+" "/VCO\_OUT-" "/I19/Vin+" 0 '1)))))}\\
         \hline
         Drain Efficiency for PA& \texttt{ymax(((- (pvi('pss "/net7" "/net4" "/I19/Vout+" 0 '(1)) / (- pvi('pss "/net8" "/gnd!" "/V1/PLUS" 0 '0)))) * 100))}\\
         \hline
         PAE for PA& \texttt{ymax((- ((100.0 * harmonic((spectralPower
         (i("/I19/Vout+" ?result "pss\_fd") (v("/net7" ?result "pss\_fd") - v("/net4" ?result "pss\_fd"))) + spectralPower(i("/I19/Vin+" ?result "pss\_fd") (v("/VCO\_OUT+" ?result "pss\_fd") - v("/VCO\_OUT-" ?result "pss\_fd")))) '(1))) / (- harmonic(spectralPower(i("/V1/PLUS" ?result "pss\_fd") (v("/net8" ?result "pss\_fd") - 0.0)) '(0))))))}\\
         \hline
    \end{tabular}
    }
    \caption{Ocean scripts implementation for different analysis of system in transmitter.}
    \label{table:Tx}
\end{table}

\chapter{A 28-GHz Receiver Design}
In order to accelerate the commercialization of high-definition video and self-driving vehicles, the mmWave frequency bands have been the great attraction over the past several decades. In addition to the increasing market needs, more advanced CMOS processes provide a feasible solution to establish the integrated circuit at this relatively high frequency, which leads to the active research area in both industry and academia \cite{razavi_architectures_1998}. A successful communication process relies on many aspects and components, such as frequency planning, transmitter design, and receiver design. Despite the fact that these devices should be optimized concurrently, the design of the dedicated receiver is worth more consideration from the perspective of signal environment and system adaptability \cite{pengfei_zhang_cmos_2005}. With the evolution of wireless standards and fabrication technology, it is becoming more important to design the mmWave receiver with low power dissipation and reasonable cost, regulated by stringent requirements.  In that regard, this section reveals a receiver architecture operating at 28 GHz and its typical performance specifications. The relevant ocean scripts implementation is also included to pave the way for the ML assisted receiver design presented in Chapter 5.

\section{Receiver Architecture}
Owing to its great tolerance and enhanced selectivity for strong interference, the heterodyne receiver manifests itself as a prevalent architecture in today’s commercial mmWave applications. The word “heterodyne” means that the Local Oscillation (LO) frequency is different from the input frequency when a mixer is applied to conduct the downconversion from Radio Frequency (RF) to Intermediate Frequency (IF) \cite{gupta_low-power_2023}. In this work, a standard frequency conversion chain is established on the receiver side by integrating the Low-Noise Amplifier (LNA) with a mixer and cascode amplifier, which can be regarded as the target system whose design process will be optimized by the ML approach later. Beginning with the receiver path, the LNA is first deployed to amplify the weak input signal from the receiving antenna without introducing excessive noise into the system. After that, the mixer is involved to carry out frequency translation as a result to generate the IF output. This output signal is then boosted by the cascode amplifier served as an IF amplifier, thereby preparing the signal for further processing. The diagram of this target system is depicted in Figure 4.1. In terms of the topology used in each individual block, the cascode LNA with an inductive degeneration, serving as the series feedback as shown in Figure 4.2 and 4.3, is deployed to provide substantial power gain while maintaining relatively low noise across a wide bandwidth \cite{molavi_wideband_2005}. As illustrated in Figure 4.4 and 4.5, the double-balanced active mixer implemented by a Gilbert cell topology is responsible for translating the frequency with the conversion gain \cite{gilbert_precise_1968}. Connecting to the output of the mixer, a low-pass filter (LPF) is implemented by a resistor and capacitor to eliminate the undesired high-frequency components. Lastly, the differential cascode amplifier is utilized to further amplify the signal with a high gain and enhanced bandwidth, which is presented in Figure 4.6 and 4.7 \cite{monsurro_09v_2009}.

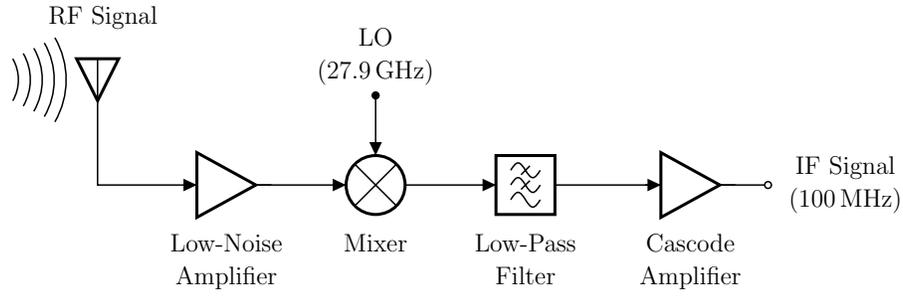
\begin{figure}[!htb]
    \centering
        \begin{circuitikz}[line width=0.2mm,scale=0.8, transform shape]
        \draw (0,0) node[rxantenna,xscale=-1](rxantenna){};
        \draw ([xshift=-0.6cm]rxantenna.north) node[above=10pt]{\text{RF Signal}};

        \draw (rxantenna.center) to[amp,name=LNA] ++(1.5,0);
        \draw (LNA.west) node[inputarrow,scale=1.5]{};
        \draw (LNA.south) node[below=5pt] {\parbox{2cm}{\centering $\text{Low-Noise}$ \\ \vspace{1pt} $\text{Amplifier}$}};
    
        \draw ([xshift=1.5cm]LNA.east) node[mixer,anchor=west](Mixer){}; 
        \draw (LNA.east) -- (Mixer.west);
        \draw (Mixer.west) node[inputarrow, scale=1.5]{};
        \draw (Mixer.north) node[inputarrow,rotate=-90,scale=1.5]{};
        \draw (Mixer.south) node[below=5pt]{$\text{Mixer}$};

        \draw (Mixer.north) to[short,-*]($(Mixer.north)+(0,1)$) node[above=1pt] {\parbox{2cm}{\centering $\text{LO}$ \\ \vspace{1pt} $\text{(27.9\,GHz)}$}};
        
        \draw ([xshift=2cm]Mixer.east) node[lowpassshape](LPF){};
        \draw (LPF.west) node[inputarrow, scale=1.5]{};
        \draw (LPF.south) node[below=5pt]{\parbox{2cm}{\centering $\text{Low-Pass}$ \\ \vspace{1pt} $\text{Filter}$}};
        \draw (Mixer.east) -- (LPF.west);

        \draw ([xshift=1.5cm]LPF.east) to[amp,name=IFAmp] ++(1.5,0);
        \draw (IFAmp.west) node[inputarrow, scale=1.5]{};
        \draw (IFAmp.south) node[below=5pt]{\parbox{2cm}{\centering $\text{Cascode}$ \\ \vspace{1pt} $\text{Amplifier}$}};
        \draw (LPF.east) -- (IFAmp.west);
        \draw (IFAmp.east) to[short, -o] ($(IFAmp.east)+(0.8,0)$) node[right=4pt] {\parbox{2cm}{\centering $\text{IF Signal}$ \\ \vspace{1pt} $\text{(100\,MHz)}$}};
    \end{circuitikz}
    \caption{A 28\,GHz heterodyne architecture comprising LNA, mixer and cascode amplifier.}
    \label{fig:Rx}
\end{figure}
\begin{figure}[t]
    \centering
    \vspace{20pt}
    \begin{circuitikz}[line width=0.1mm,scale=0.8, transform shape]
        \ctikzset{tripoles/mos style/arrows}
        \ctikzset{capacitors/scale=0.4}
        \ctikzset{resistors/scale=0.5}
        \ctikzset{inductors/scale=0.8, inductors/coils=4}
        \ctikzset{transistors/scale=0.8}
        \ctikzset{grounds/scale=0.8}

        \draw (0,0) node[nmos,anchor=G](M1){$\text{W}_\text{N1}$};

        \draw (M1.D)--++(0,0.3) node[nmos,anchor=S](M2){$\text{W}_\text{N2}$};
        \draw (M2.G)--++(-0.5,0);
        \draw ([xshift=-0.5cm]M2.G)--++(0,2.2);
        \draw (M2.D) --++(0.8,0);

        \draw (M2.D) to[L,l=$\text{L}_\text{d}$,name=Ld,mirror] ++(0,1.6);
        \draw (M1.G) to[L,l=$\text{L}_\text{g}$,name=Lg,mirror] ++(-1.4,0);
        \draw (M1.S) to[L,l=$\text{L}_\text{s}$,name=Ls] ++(0,-1.6);

        \draw ([xshift=0.8cm]M2.D) to[C,l_=$\text{C}_\text{1}$,name=C1] ++(1.6,0);
        \draw ([xshift=0.6cm]C1.east) to[C,name=C3] ++(0,-1.4);
        \draw ([xshift=0.75cm]C1.east) to[short,-o] ($(C1.east)+(1,0)$) node[right] {$\text{V}_\text{out}$};
        \draw ([xshift=-0.8cm]Lg.west) to[C,l=$\text{C}_\text{2}$,name=C2] ++(-1.4,0);
        \draw ([xshift=-0.75cm]C2.west) to[short,-o] ($(C2.west)-(0.9,0)$) node[left] {$\text{V}_\text{in}$};

        \draw ([xshift=0.8cm]M2.D) to[R,name=R3] ++(0,1.6);
        \draw ([xshift=-0.3cm]Lg.east) to[R,name=R2] ++(0,1.7);

        \draw ([xshift=-0.2cm,yshift=0.56cm]R2.east) node[nmos,anchor=G,xscale=-1](M3){};
        \draw (M3.D) to[R,name=R1] ++(0,1.45);
        \draw (M3.G) --++(0.2,0);
        \draw ([xshift=0.2cm]M3.G) --++(0,0.7);
        \draw ([xshift=0.2cm,yshift=0.7cm]M3.G) --++(-0.98,0);

        \draw [line width=0.5mm]([yshift=0.8cm,xshift=-4cm]Ld.north)--([yshift=0.8cm,xshift=2.5cm]Ld.north) node[above,near end]{$\textbf{V}_\textbf{DD}$};

        \draw (M3.S) node[ground](GND){};
        \draw ([yshift=-0.4cm]Ls.east) node[ground](GND){};
        \draw ([yshift=-0.6cm]C3.east) node[ground](GND){};
        
    \end{circuitikz}
    \caption{Cascode LNA schematic with an inductive degeneration.}
    \label{fig:LNA}
    \vspace{30pt}
\end{figure}
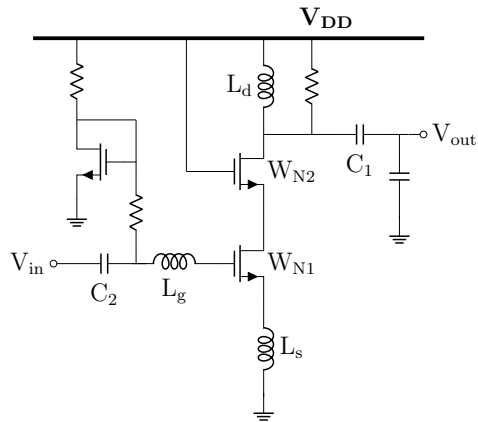
\begin{figure}[!htb]
    \centering
    \includegraphics[width=0.8\textwidth]{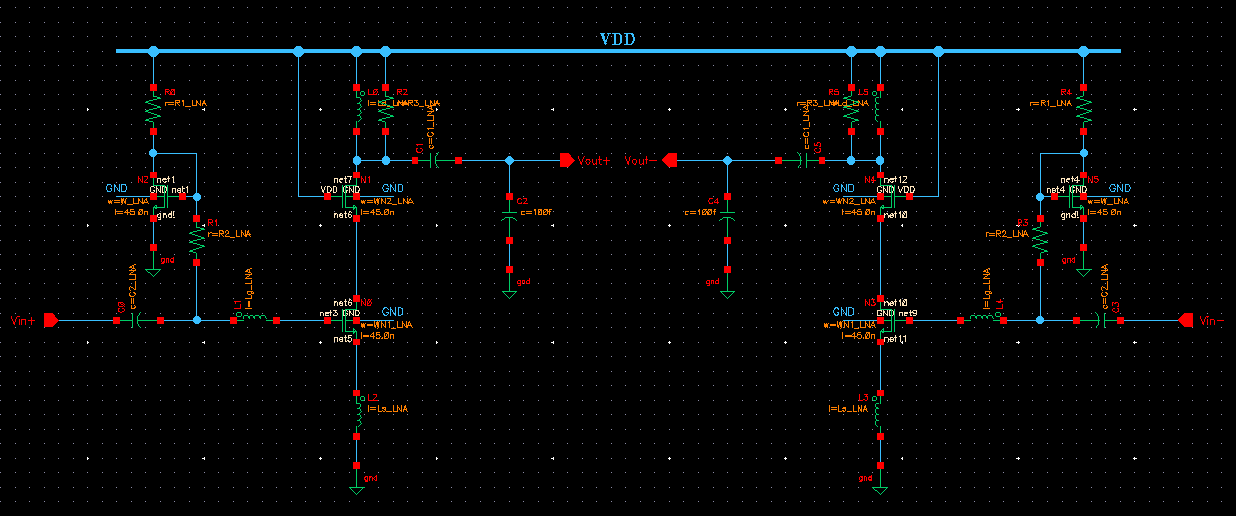}
    \caption{LNA implementation with differential configuration using NCSU 45\,nm PDK in Cadence Virtuoso}
    \label{fig:LNA_Cadence}
    \vspace{20pt}
\end{figure}
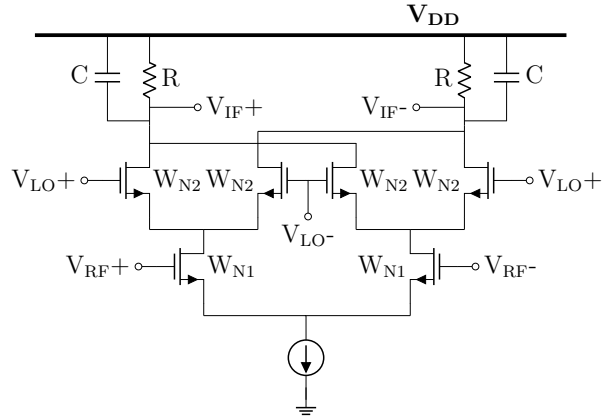
\begin{figure}[!htb]
    \centering
    \begin{circuitikz}[line width=0.1mm,scale=0.8, transform shape, american]
        \ctikzset{tripoles/mos style/arrows}
        \ctikzset{capacitors/scale=0.5}
        \ctikzset{resistors/scale=0.5}
        \ctikzset{inductors/scale=0.8, inductors/coils=4}
        \ctikzset{transistors/scale=0.8}
        \ctikzset{grounds/scale=0.8}
        \ctikzset{sources/scale=0.7}

        \draw (0,0) node[nmos,anchor=G](M1){$\text{W}_\text{N1}$};
        \draw (M1.G) to[short,-o] ($(M1.G)-(0.3,0)$) node[left] {$\text{V}_\text{RF}\text{+}$};

        \draw (5,0) node[nmos,anchor=G,xscale=-1](M2){\ctikzflipx{$\text{W}_\text{N1}$}};
        \draw (M2.G) to[short,-o] ($(M2.G)+(0.3,0)$) node[right] {$\text{V}_\text{RF}\text{-}$};

        \draw (M1.D) -- ([xshift=-0.9cm]M1.D);
        \draw ([xshift=-0.9cm]M1.D) --++(0,0.2) node[nmos,anchor=S](M3){$\text{W}_\text{N2}$};
        \draw (M3.G) to[short,-o] ($(M3.G)-(0.3,0)$) node[left] {$\text{V}_\text{LO}\text{+}$};
        \draw ([yshift=0.6cm]M3.D) to[short,-o] ($(M3.D)+(0.8,0.6)$) node[right] {$\text{V}_\text{IF}\text{+}$};

        \draw (M1.D) -- ([xshift=0.9cm]M1.D);
        \draw ([xshift=0.9cm]M1.D) --++(0,0.2) node[nmos,anchor=S,xscale=-1](M4){\ctikzflipx{$\text{W}_\text{N2}$}};
        \draw ([xshift=0.05cm]M4.G) to[short,-o] ($(M4.G)+(0.05,-0.6)$) node[below] {$\text{V}_\text{LO}\text{-}$};

        \draw (M2.D) -- ([xshift=-0.9cm]M2.D);
        \draw ([xshift=-0.9cm]M2.D) --++(0,0.2) node[nmos,anchor=S](M5){$\text{W}_\text{N2}$};
        \draw (M4.G) -- (M5.G);
        \draw (M3.D) -- (M5.D);

        \draw (M2.D) -- ([xshift=0.9cm]M2.D);
        \draw ([xshift=0.9cm]M2.D) --++(0,0.2) node[nmos,anchor=S,xscale=-1](M6){\ctikzflipx{$\text{W}_\text{N2}$}};
        \draw (M4.D) -- ([yshift=0.2cm]M4.D);
        \draw ([yshift=0.2cm]M4.D) -- ([yshift=0.2cm]M6.D);
        \draw (M6.G) to[short,-o] ($(M6.G)+(0.3,0)$) node[right] {$\text{V}_\text{LO}\text{+}$};
        \draw ([yshift=0.6cm]M6.D) to[short,-o] ($(M6.D)+(-0.8,0.6)$) node[left] {$\text{V}_\text{IF}\text{-}$};

        \draw (M3.D) -- ([yshift=0.3cm]M3.D);
        \draw ([yshift=0.3cm]M3.D) to[R,l_=$\text{R}$,name=R1] ++(0,1.5);
        \draw (M6.D) -- ([yshift=0.3cm]M6.D);
        \draw ([yshift=0.3cm]M6.D) to[R,l=$\text{R}$,name=R2] ++(0,1.5);

        \draw ([xshift=-0.7cm,yshift=-0.4cm]R1.west) to[C,l=$\text{C}$,name=C1] ++(0,1.45);
        \draw ([yshift=-0.65cm]C1.west) --([xshift=0.7cm,yshift=-0.65cm]C1.west);
        \draw ([xshift=0.7cm,yshift=-0.4cm]R2.west) to[C,l_=$\text{C}$,name=C2] ++(0,1.45);
        \draw ([yshift=-0.65cm]C2.west) --([xshift=-0.7cm,yshift=-0.65cm]C2.west);

        \draw [line width=0.5mm]([yshift=0.75cm,xshift=-1.8cm]R1.north)--([yshift=0.75cm,xshift=1.8cm]R2.north) node[above,near end]{$\textbf{V}_\textbf{DD}$};

        \draw (M1.S) -- ([yshift=-0.2cm]M1.S);
        \draw ([yshift=-0.2cm]M1.S) -- ([xshift=1.7cm, yshift=-0.2cm]M1.S);
        \draw (M2.S) -- ([yshift=-0.2cm]M2.S);
        \draw ([yshift=-0.2cm]M2.S) -- ([xshift=-1.75cm, yshift=-0.2cm]M2.S);
        \draw ([xshift=1.7cm, yshift=-0.2cm]M1.S) to[I,name=I1] ++(0,-1.4);

        \draw ([yshift=-0.2cm]I1.right) node[ground](GND){};
        
    \end{circuitikz}
    \caption{Double-balanced Gilbert cell mixer schematic.}
    \label{fig:Mixer}
    \vspace{20pt}
\end{figure}
\begin{figure}[!htb]
    \centering
    \includegraphics[width=0.4\textwidth]{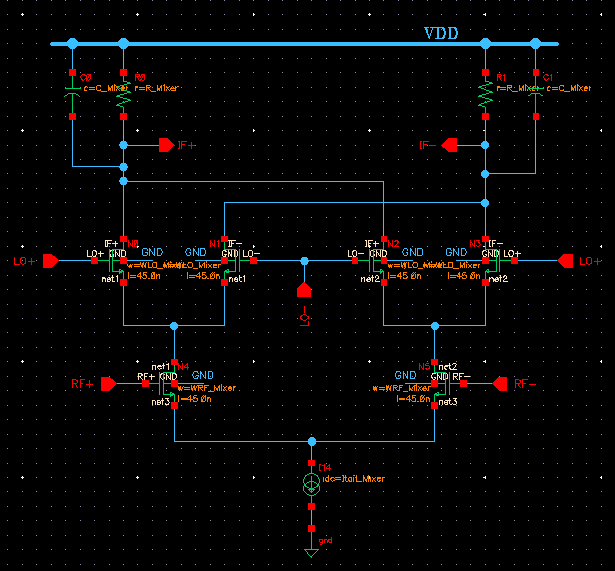}
    \caption{Mixer implementation using NCSU 45\,nm PDK in Cadence Virtuoso}
    \label{fig:Mixer_Cadence}
    \vspace{20pt}
\end{figure}
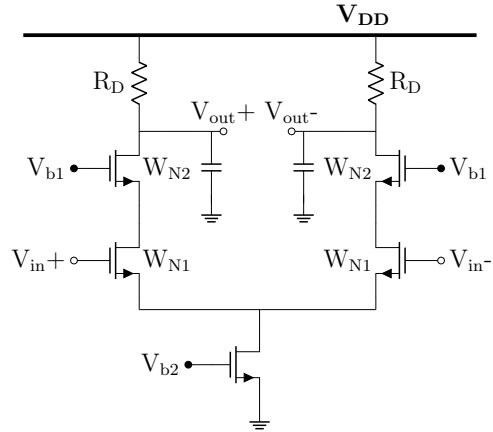
\begin{figure}[!htb]
    \centering
    \begin{circuitikz}[line width=0.1mm,scale=0.8, transform shape]
        \ctikzset{tripoles/mos style/arrows}
        \ctikzset{capacitors/scale=0.4}
        \ctikzset{resistors/scale=0.6}
        \ctikzset{transistors/scale=0.8}
        \ctikzset{grounds/scale=0.8}
        
        \draw (0,0) node [nmos, anchor = G](M1){$\text{W}_\text{N1}$};
        \draw (M1.G) to[short, -o] ($(M1.G)-(0.3,0)$) node[left] {$\text{V}_\text{in}\text{+}$};
        
        \draw (5.5,0) node [nmos, anchor = G, xscale=-1](M3){\ctikzflipx{$\text{W}_\text{N1}$}};
        \draw (M3.G) to[short, -o] ($(M3.G)+(0.3,0)$) node[right] {$\text{V}_\text{in}\text{-}$};
        
        \draw (M1.D) -- ++(0,0.3) node[nmos, anchor=S, xscale=1](M2) {$\text{W}_\text{N2}$};
        \draw (M2.G) to[short, -*] ($(M2.G)-(0.3,0)$) node[left] {$\text{V}_\text{b1}$};
        \draw (M2.D) to[short, -o] ($(M2.D)+(1.4,0)$) node[above] {$\text{V}_\text{out}\text{+}$};
        \draw ([xshift=1.2cm]M2.D) to[C, name=C1] ++(0,-1.2);
        \draw ([yshift=-0.4cm]C1.east) node[ground](GND){};
        
        \draw (M3.D) -- ++(0,0.3) node[nmos, anchor=S, xscale=-1](M4) {\ctikzflipx{$\text{W}_\text{N2}$}};
        \draw (M4.G) to[short, -*] ($(M4.G)+(0.3,0)$) node[right] {$\text{V}_\text{b1}$};
        \draw (M4.D) to[short, -o] ($(M4.D)-(1.4,0)$) node[above] {$\text{V}_\text{out}\text{-}$};
        \draw ([xshift=-1.2cm]M4.D) to[C, name=C2] ++(0,-1.2);
        \draw ([yshift=-0.4cm]C2.east) node[ground](GND){};
        
        \draw (M2.D) to[R,l=$\text{R}_\text{D}$,name=R1] ++(0,1.6);
        \draw (M4.D) to[R,l_=$\text{R}_\text{D}$,name=R2] ++(0,1.6);
        
        \draw [line width=0.5mm] ([yshift=0.8cm, xshift=-1.8cm]R1.north) -- ([yshift=0.8cm,xshift=1.8cm]R2.north) node[above, near end] {$\textbf{V}_\textbf{DD}$};
        
        \draw (M1.S) -- ++(0,-0.2);
        \draw (M3.S) -- ++(0,-0.2);
        \draw ([yshift=-0.2cm]M1.S) -- ([yshift=-0.2cm]M3.S);
        \draw ([xshift=2cm, yshift=-0.5cm]M1.S) node [nmos, anchor = D](M5){};
        \draw ([xshift=2cm, yshift=-0.2cm]M1.S) -- (M5.D);
        \draw (M5.G) to[short, -*] ($(M5.G)-(0.4,0)$) node[left] {$\text{V}_\text{b2}$};
        
        \draw (M5.S) node[ground](GND){};
        
    \end{circuitikz}
    \caption{Differential cascode amplifier schematic.}
    \label{fig:Cascode}
\end{figure}
\begin{figure}[!htb]
    \centering
    \vspace{20pt}
    \includegraphics[width=0.6\textwidth]{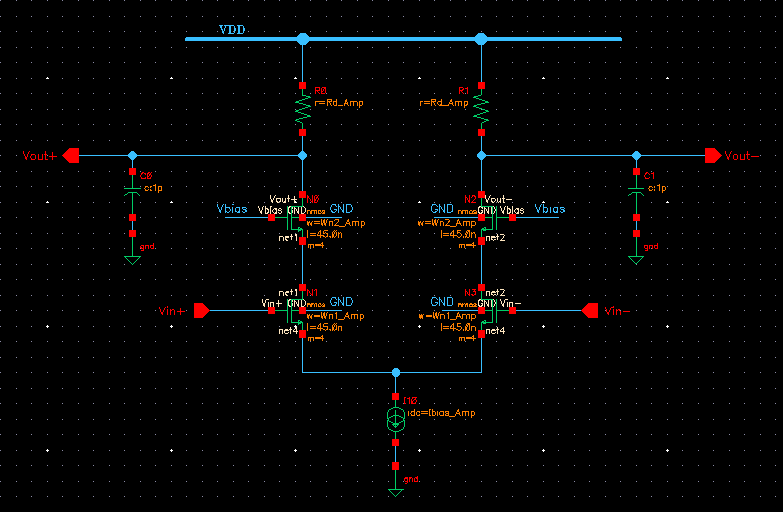}
    \caption{Cascode Amplifier implementation using NCSU 45\,nm PDK in Cadence Virtuoso}
    \label{fig:Cascode_Cadence}
    \vspace{20pt}
\end{figure}

\section{Performance Specifications Analysis}
\subsection{Individual Performance}
This part mainly relies on detailed performance specifications of the individual blocks in the receiver systems. These metrics manifest themselves as critical considerations both in the conventional circuit design process and in the new method using the ML approach demonstrated in this work.

As the first active building block in the receiver front-end, the primary goal of the LNA is to amplify the signal from the receiving antenna with significant gain and relatively low noise. In that respect, the specification related to power gain should first be taken into consideration. In order to clearly indicate this performance, the transducer power gain is used in the LNA to obtain the ratio between the power delivered to the load and the power available from the source, \text{\(\Gamma_{\text{S}}\) and \(\Gamma_{\text{L}}\)} represent the reflection coefficient on the source and load terminals (Eq.\,(4.1)):
\begin{equation}
    G_T = \frac{(1 - |\Gamma_S|^2) |S_{21}|^2}{|1 - S_{11} \Gamma_S|^2} \times \frac{1 - |\Gamma_L|^2}{|1 - S_{22} \Gamma_L|^2}
\end{equation}
In conjunction with the power gain, noise figure is another important metric to evaluate the ability to suppress the excessive noise in LNA, which is quantified by the ratio of signal-to-noise ratio between the input and output (Eq.\,(4.2)):
\begin{equation}
    NF = \frac{SNR_{in}}{SNR_{out}} = \frac{S_i/N_i}{S_o/N_o} = \frac{S_i/ N_i}{GS_i / (N_a + GN_i)} = 1 + \frac{N_a}{GN_i}
\end{equation}
Since the noise comes from the random process and it cannot be completely eliminated, the value of noise figure is always larger than 1. In addition, an essential entity called the scattering parameters is implemented to determine the relationships between the inputs and outputs in multi-port system. Specifically, the input port voltage reflection coefficient ($\text{S}_\text{11}$) is worth further consideration in the LNA design at 28 GHz since it can be employed to derive other important specifications, such as input impedance and return loss. 

In terms of mixer performance metrics, the primary specification should be conversion gain due to its active configuration. The term “conversion” indicates that the value of this entity comes from the ratio between the output voltage in IF band and the input voltage in RF band (Eq.\,(4.3)):
\begin{equation}
    \frac{V_{IF}}{V_{RF}} = \frac{2}{\pi} g_{m} R_D
\end{equation}
Besides the conversion gain, voltage swing can be also regarded as an important metric in the mixer design to evaluate the ability to tolerate the large input signal without clipping or compression, thereby examining the stability. 

Concerning the relatively simple topology and straightforward relationship between the circuit parameters and relevant performance specifications, the voltage gain is the only selected metric to signify the capability of amplification in the cascode amplifier design, which is calculated based on the ratio between the output voltage and the input voltage. 

\subsection{System Performance}
For the RF receiver design, the performance named sensitivity manifests itself as a critical specification due to the hostile wireless communication environment, in which the minimum signal level for detecting the inputs with tolerable quality is derived from the noise figure and the output signal-to-noise ratio, as shown in Eq.\,(4.4):
\begin{equation}
    S_{sen}|_{dBm} = -174 + NF_{dB} + 10 \log \Delta f + SNR_{out}
\end{equation}
In addition to the sensitivity, the maximum receiving signal power is constrained by 1 dB compression point to signify the high end for the linear operating range when the real output power decreases by 1 dB from the theoretical value. The above two metrics constitute an essential specification called the dynamic range to identify the desired range for successfully detecting the receiving signal with acceptable noise and distortion. In this work, the power consumption and voltage gain are first investigated to examine the energy efficiency and dynamic range of the system comprising the LNA, mixer, and cascode amplifier on receiver side. Moreover, the noise figure is applied to measure the sensitivity performance of the target system, which is based on the Friis’ equation (Eq.\,(4.5)):
\begin{equation}
    NF_{Total} = NF_1 + \frac{NF_2 - 1}{G_{A1}} + \cdots + \frac{NF_N - 1}{G_{A1} \times G_{A2} \times \cdots \times G_{A_{N-1}}}
\end{equation}

\section{Ocean Scripts Implementation}
The aforementioned contents present the necessity and benefits of utilizing ocean scripts to facilitate repetitive simulations and parametric analysis, it is becoming increasingly more important to conduct the simulation of the complicated system with relatively low time cost and bridge the gap between the Cadence simulator and Python environment. 

After integrating the different analysis settings with the corresponding circuit netlists, the ocean file is run from the command line to collect the performance of the target receiver system. Specifically, dc analysis is used to obtain system power consumption with 1.2 V supply voltage. As the indispensable approaches to simulate the periodically-varying circuits, some processes adhered to the periodic steady-state (PSS) analysis are employed to capture the system performance including the voltage gain and noise figure through the Periodic AC (PAC) analysis and Periodic S Parameters (PSP) analysis respectively. In light of the individual block specifications, the Quasi-Periodic Steady State (QPSS) analysis equipped with the shooting engine is adopted to access the gain capability of the system on the receiver side, comprising the power gain of LNA, the conversion gain of mixer, and the voltage gain of cascode amplifier. In addition, SP analysis is leveraged to quantify the noise figure and $\text{S}_\text{11}$ of LNA.  For the voltage swing of the mixer, it can be determined by the transient analysis. The schematic of the target system in Cadence and the relevant performance metrics with the ocean scripts implementation are presented in Figure 4.8 and Table 4.1. 
\begin{figure}[!htb]
    \centering
    \includegraphics[width=1\textwidth]{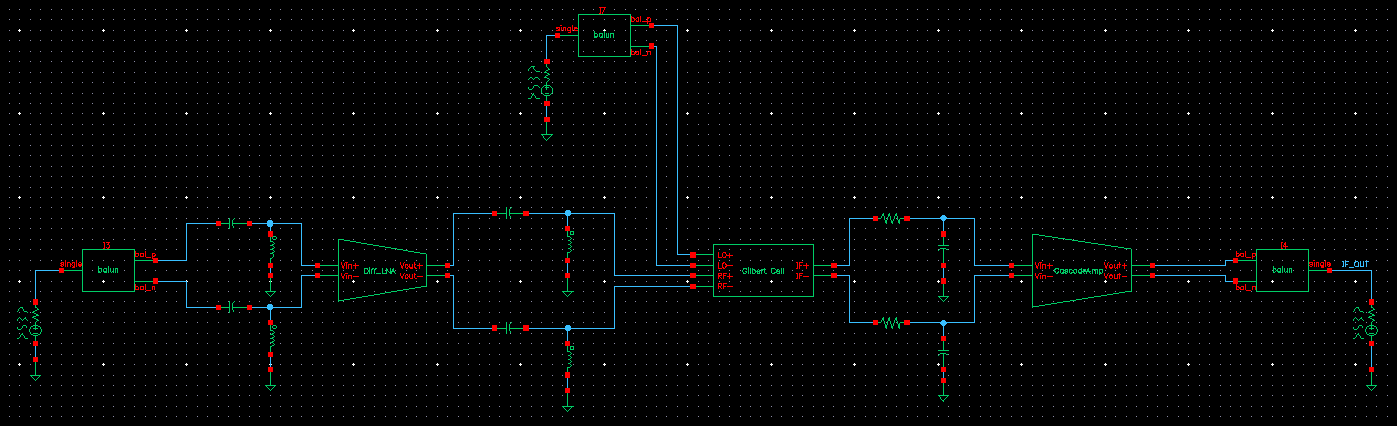}
    \caption{Schematic of target system on receiver side in Cadence Virtuoso.}
    \label{fig:Rx_Cadence}
    \vspace{10pt}
\end{figure}
\begin{table}[t]
    \centering
    \renewcommand{\arraystretch}{1.5}
    {\small \begin{tabular}{m{0.4\textwidth}m{0.5\textwidth}}
        \hline
        \textbf{Performance Metrics} & \textbf{Ocean Scripts}\\
        \hline\hline
         Power Consumption& \texttt{getData(":pwr" ?result "dcOp")}\\
         \hline
         Voltage Gain& \texttt{ymax(db(vh('pac "/IF\_OUT" '-1)))}\\
         \hline
         Noise Figure& \texttt{ymin(getData("NF" ?result "psp"))}\\
         \hline
         Power Gain for LNA& \texttt{ymax(db10((pvi('qpss "/LNA\_OUT+" "/LNA\_OUT-" "/I0/Vout+" 0) / (- pvi('qpss "/LNA\_IN+" "/LNA\_IN-" "/I0/Vin+" 0 '(0 1))))))}\\
         \hline
         Noise Figure for LNA& \texttt{ymin(db10(getData("F" ?result "sp\_noise")))}\\
         \hline
         $\text{S}_\text{11}$ for LNA& \texttt{ymin(db(spm('sp 1 1)))}\\
         \hline
         Conversion Gain for Mixer& \texttt{ymax(db10((pvi('qpss "/Mixer\_OUT+" "/Mixer\_OUT-" "/I1/IF+" 0) / (- pvi('qpss "/Mixer\_IN+" "/Mixer\_IN-" "/I1/RF+" 0 '(0 1))))))}\\
         \hline
         Voltage Swing for Mixer& \texttt{(ymax((vtime('tran "/Mixer\_OUT+") - vtime('tran "/Mixer\_OUT-") - (v("/Mixer\_OUT+" ?result "dcOp") - v("/Mixer\_OUT-" ?result "dcOp")))) - ymin((vtime('tran "/Mixer\_OUT+") - vtime('tran "/Mixer\_OUT-") - (v("/Mixer\_OUT+" ?result "dcOp") - v("/Mixer\_OUT-" ?result "dcOp"))))) / 2}\\
         \hline
         Voltage Gain for Cascode Amplifier& \texttt{ymax(db((vh('qpss "/Amp\_OUT+") / harmonic(vh('qpss "/Amp\_IN+") '((-1 1))))))}\\
         \hline
    \end{tabular}
    }
    \caption{Ocean scripts implementation for different analysis of system in receiver.}
    \label{table:Rx}
\end{table}
\chapter{Machine Learning Based Transceiver Circuits Design}
The previous contents have elaborated the architecture of two systems on both the transmitter and receiver sides whose operating frequency is 28 GHz. Since these systems include heterogeneous blocks, such as the LNA, mixer, and cascode amplifier in the receiver architecture, the direction of optimizing their performance cannot be obtained in a straightforward way. In addition, more sophisticated CMOS processes and relatively intricate trade-offs lead to other impediments to the mmWave circuit design. In that regard, this section primarily concentrates on the ML approaches consisting of the problem statement, benchmark data collection, and model evaluation leveraging the conventional and large-scale algorithms, which are employed to accelerate the design process of mmWave circuits.

\section{Problem Statement}
Typically, the circuit design starts with the selection of the appropriate topology confined by the desired performance, such as the power consumption and the voltage gain. Following that, the key circuit parameters in the chosen topology are identified and determined by experienced circuit designers leveraging their knowledge to find the approximate solution and then conduct iteration for the values of diversified parameters until the required specifications are fulfilled. In contrast, ML-based circuit design carries out a reverse process, as formulated in Eq.\,(5.1):
\begin{equation}
    \bm{x} = \mathcal{M}(\bm{y}),
\end{equation}
In particular, $\mathcal{M}$ represents the dedicated ML model to predict the output $\bm{x}$ that entails a set of circuit parameters given the performance specification vectors denoted by $\bm{y}$. This approach is deployed to directly learn the relationship between the desired performance and circuit parameters, thereby accelerating the design process rather than implementing the time-consuming iterations. 

\section{Data Collection}
For the two target systems with five different individual blocks, the corresponding circuit schematics are first created in Cadence Virtuoso using NCSU 45 nm Processing Design Kit (PDK). Then, the initial circuit parameters are set to reasonable values developed by the prior knowledge of the circuit designer as a result to provide a reliable starting point for ML model training later. For each parameter, the value is swept with a small step size in a certain variation range, and the related performance metrics are obtained from the simulation results shown in Cadence spectre simulator. Based on the complexity of each circuit block, a different number of parameters are selected during the process of simulation. In addition, the length of each transistor is fixed at 45\,nm to simplify the design space and mitigate short-channel effects. In order to facilitate the training process, the circuit parameters are merged with the corresponding specifications to establish the dataset that will be further split by the training and testing components. The entire data collection procedure is illustrated in Figure 5.1, and the sweeping range of circuit parameters presented in the created dataset is listed in Table 5.1.

\input{Figures/fig_dataset}
\captionsetup[table]{justification=centering}
\begin{table}[!htb]
    \centering
    \noindent\resizebox{\textwidth}{!}{
       \begin{tabular}{l|l|cc}
        \toprule
         \textbf{System-Level Circuits} & \textbf{Individual Block} & \textbf{Parameter} & \textbf{Sweeping Range} \\ \midrule
         \multirow{12}{*}{\makecell[l]{Transmitter system \\ \textbf{specs}: \\ dc power $|$ bw $|$ output power \\ 
         $|$ voltage swing}} & \multirow{6}{*}{\makecell[l]{Voltage-Controlled Oscillator (VCO) \\ \textbf{specs}: \\ phase noise $|$ tuning range}}& $\text{C}$ &  $\text{50:50:150 (fF)}$ \\
         & & $\text{L}$ & $\text{60:60:180 (pH)}$ \\ 
         & & $\text{R}_{\text{p}}$ & $\text{300:100:500 (\(\Omega\))}$ \\ 
         & & $\text{W}_{\text{N1}}$ & $\text{7.5:2.5:12.5 (\(\upmu\)m)}$ \\ 
         & & $\text{W}_{\text{N2}}$ & $\text{187.5:12.5:212.5 (\(\upmu\)m)}$ \\ 
         & & $\text{W}_{\text{var}}$ & $\text{70:10:90 (\(\upmu\)m)}$ \\ 
         \cmidrule{2-4}
         & \multirow{6}{*}{\makecell[l]{Power Amplifier (PA)  \\ \textbf{specs}: \\ power gain $|$ drain efficiency $|$ PAE}} & $\text{L}_{\text{ip}}$ & $\text{175:175:350 (pH)}$ \\
         & & $\text{L}_{\text{is}}$ & $\text{60:60:120 (pH)}$ \\ 
         & & $\text{L}_{\text{op}}$ & $\text{360:353:713 (pH)}$ \\ 
         & & $\text{L}_{\text{os}}$ & $\text{45:45:90 (pH)}$ \\
         & & $\text{W}_{\text{N1}}$ & $\text{22:5:32 (\(\upmu\)m)}$ \\ 
         & & $\text{W}_{\text{N2}}$ & $\text{16:5:26 (\(\upmu\)m)}$ \\ \midrule

         \multirow{14}{*}{\makecell[l]{Receiver system  \\ \textbf{specs}: \\ dc power $|$ gain $|$ noise figure}}& \multirow{7}{*}{\makecell[l]{Low-Noise Amplifier (LNA)  \\ \textbf{specs}: \\ power gain $|$ $\text{S}_{\text{11}}$ $|$ noise figure}}& $\text{C}_{\text{1}}$ & $\text{130:50:180 (fF)}$ \\
         & & $\text{C}_{\text{2}}$ & $\text{170:50:220 (fF)}$ \\ 
         & & $\text{L}_{\text{d}}$ & $\text{180:50:230 (pH)}$ \\ 
         & & $\text{L}_{\text{g}}$ & $\text{850:100:950 (pH)}$ \\ 
         & & $\text{L}_{\text{s}}$ & $\text{80:10:90 (pH)}$ \\ 
         & & $\text{W}_{\text{N1}}$ & $\text{20:3:26 (\(\upmu\)m)}$ \\ 
         & & $\text{W}_{\text{N2}}$ & $\text{37.5:2.5:42.5 (\(\upmu\)m)}$ \\ 
         \cmidrule{2-4}
         & \multirow{4}{*}{\makecell[l]{Mixer \\ \textbf{specs}: \\ voltage swing $|$ conversion gain}} & $\text{C}$ & $\text{1:0.1:1.1 (pF)}$ \\
         & & $\text{R}$ & $\text{400:100:500 (\(\Omega\))}$ \\ 
         & & $\text{W}_{\text{N1}}$ & $\text{14:2:18 (\(\upmu\)m)}$ \\ 
         & & $\text{W}_{\text{N2}}$ & $\text{6:2:10 (\(\upmu\)m)}$ \\
         \cmidrule{2-4}
         & \multirow{3}{*}{\makecell[l]{Cascode Amplifier \\ \textbf{specs}: \\ gain}} & $\text{R}_{\text{D}}$ & $\text{300:100:400 (\(\Omega\))}$ \\
         & & $\text{W}_{\text{N1}}$ & $\text{26:2:30 (\(\upmu\)m)}$ \\ 
         & & $\text{W}_{\text{N2}}$ & $\text{14:2:18 (\(\upmu\)m)}$ \\ 
         \bottomrule
      \end{tabular}

    }
    \caption{The chosen parameters and specifications for transceiver circuits and each individual block. The sweeping range of selected design parameters is written in the form of [begin, increment, end].}
    \label{tab:circuitparam:complex}
\end{table}

\section{Model Evaluation}
With the dataset collected from the two mmWave systems following the procedure illustrated above, some conventional and modern large-scale ML algorithms are employed to determine the relationship between the input specifications and the output parameters. In particular, the Random Forest (RF), Support Vector Regression (SVR), Multi-layer Perceptron (MLP), and Transformer are utilized to develop the predictions for diversified circuit parameters based on desired performance metrics. To cope with high nonlinearity and intricate trade-offs of mmWave system design, the most nonlinear kernel named rbf is selected for SVR model, and a rectified linear unit (ReLU) activation function is utilized in MLP model. The detailed characteristics for each model are itemized in Table 5.2 and the evaluation of these models is also presented using the mean-square loss as an objective function. As shown in Eq.\,(5.2), the difference between the performance based on the predicted parameters $\hat{\bm{y}}$ and the desired counterpart $\bm{y}$ is formulated by means of individual error in each specification and the aggregated error.
\begin{equation}
    \text{individual error}: err_i = \| \bm{y}_i - \hat{\bm{y}}_i \| / \bm{y}_i \quad \text{aggregated error}: err = \| \bm{y} - \hat{\bm{y}} \| / \bm{y}
\end{equation}

\captionsetup[table]{justification=centering}
\begin{table}[!htb]
    \centering
    \vspace{20pt}
    \noindent\resizebox{0.9\textwidth}{!}{
    \begin{tabular}{l|cc}
        \toprule
        \textbf{Model} & \textbf{Parameter} & \textbf{Value} \\ \midrule
        \multirow{2}{*}{\makecell[l]{Random Forest (RF)}} & $\texttt{n\_estimators}$& $\text{100}$\\
           & $\texttt{criterion}$& $\text{squared\_error (L2 Loss)}$\\ \midrule
        \multirow{2}{*}{\makecell[l]{Support Vector Regressor (SVR)}} & $\texttt{kernel}$& $\text{rbf}$\\
           & $\texttt{multi\_target\_regression\_type}$& $\text{MultiOutputRegression}$\\\midrule
        \multirow{2}{*}{\makecell[l]{Multi-layer Perception (MLP)}} & $\texttt{num\_layers}$& $\text{7}$\\
           & $\texttt{dim\_layers}$& $\text{[200, 300, 500, 500, 300, 200]}$\\ \midrule
        \multirow{6}{*}{\makecell[l]{Transformer}} & $\texttt{dim\_model}$ & $\text{200}$ \\
            & $\texttt{num\_heads}$  & $\text{2}$\\
            & $\texttt{dim\_hidden}$ & $\text{200}$\\
            & $\texttt{dropout\_p}$ & $\text{0.1}$\\
            & $\texttt{num\_encoder\_layers}$& $\text{6}$\\
            & $\texttt{activation}$& $\text{relu}$\\
         \bottomrule
    \end{tabular}

    }
    \caption{Various ML models and their selected parameters.}
    \label{tab:model}
    \vspace{10pt}
\end{table}

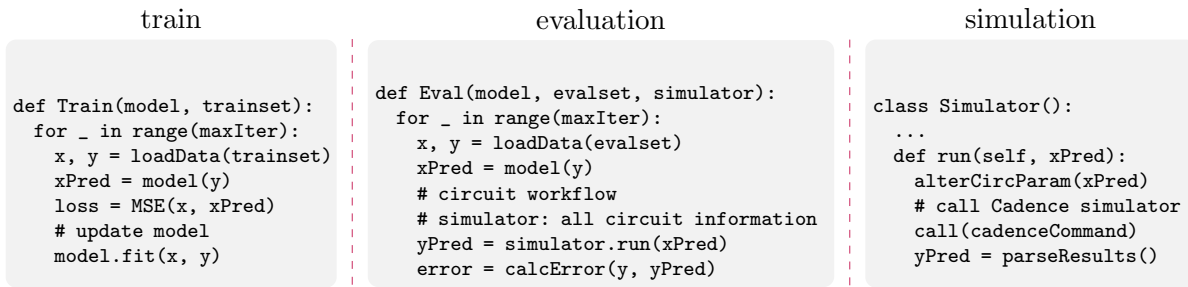
\begin{figure}[!htb]
\centering
\begin{tikzpicture}
\node[draw=none, fill=gray!10, minimum height=3.3cm, text width=4.2cm, rounded corners, label=above:{\small train}, inner sep=1mm, align=left](train){\scriptsize
\begin{verbatim}
def Train(model, trainset):
  for _ in range(maxIter):
    x, y = loadData(trainset)
    xPred = model(y)
    loss = MSE(x, xPred)
    # update model
    model.fit(x, y)
\end{verbatim}
};

\draw[-, dashed, purple, opacity=.8] ($(train.north east) + (2mm, 0mm)$) -- ($(train.south east) + (2mm, 0mm)$);

\node[right=.4cm of train, draw=none, fill=gray!10, minimum height=3.2cm, text width=6cm, rounded corners, label=above:{\small evaluation}, inner sep=1mm](test){\scriptsize
\begin{verbatim}
def Eval(model, evalset, simulator):
  for _ in range(maxIter):
    x, y = loadData(evalset)
    xPred = model(y)
    # circuit workflow
    # simulator: all circuit information
    yPred = simulator.run(xPred)
    error = calcError(y, yPred)
\end{verbatim}

};

\draw[-, dashed, purple, opacity=.8] ($(test.north east) + (2mm, 0mm)$) -- ($(test.south east) + (2mm, 0mm)$);

\node[right=.4cm of test, draw=none, fill=gray!10, minimum height=3.3cm, text width=4.2cm, rounded corners, label=above:{\small simulation}, inner sep=1mm](simulator){\scriptsize
\begin{verbatim}
class Simulator():
  ...
  def run(self, xPred):
    alterCircParam(xPred)
    # call Cadence simulator
    call(cadenceCommand)
    yPred = parseResults()
\end{verbatim}
};
\end{tikzpicture}
\caption{An end-to-end model training and evaluation pipeline.}
\label{fig:pipeline}
\vspace{10pt}
\end{figure}

In terms of Python implementation, an end-to-end model training and evaluation pipeline is adopted in this work, which facilitates the interaction between the ML workflow and the circuit design process, as depicted in Figure 5.2. By employing the trained model, the predicted parameters based on the desired specifications are obtained in the first place. Then, the simulator in Cadence Virtuoso is invoked to conduct the simulation and compare the theoretical performance with the predicted performance, thereby calculating the relative error by virtue of the mean-square loss function mentioned above.

\chapter{Conclusion and Future Works}
\section{Conclusion}
In this thesis, the background of 5G wireless communication using the mmWave frequency band is introduced comprising top-level technologies and integrated circuit application. Following that, the motivation and literature review sections indicate the necessity of establishing the automation pipeline for analog/RF circuit design. In that case, two target mmWave systems are investigated on the transmitter and receiver sides with the schematic design and the performance metrics selection, in which the chosen specifications are implemented by ocean scripts leveraging various analysis methods. Consequently, ML approaches are explored to facilitate circuit design with different models consisting of conventional and large-scale algorithms, which entails superior performance for both accuracy and efficiency.

\section{Future Works}
In this section, some possible directions of future work are specified from the perspective of circuit design and ML algorithms. In terms of circuit design, topology selection and layout establishment can be regarded as other critical tasks that need to be optimized by ML approaches attributed to time-consuming and labor-intensive processes. For the ML algorithms point of view, the reinforcement learning and graph convolutional neural network (GCN) can also be considered as the powerful candidates to optimize the design process for mmWave integrated circuits. Finally, the sampling methods leading to the enhanced data efficiency are worth more research efforts in the future, especially for more complicated system design.

\clearpage
\phantomsection
\addcontentsline{toc}{chapter}{Bibliography}

\begingroup
\setstretch{1}
\printbibliography
\endgroup

\captionsetup[figure]{list=no}
\captionsetup[table]{list=no}


\end{document}